\documentclass[a4paper,twoside,12pt,openright]{report} 
\usepackage{amsthm} 
\usepackage{bbm}
\usepackage[ruled, lined, linesnumbered, commentsnumbered, longend]{algorithm2e}
\usepackage{comment}
\usepackage[pdeec,final, backrefs]{feupphdteses}
\usepackage[utf8]{inputenc}
\usepackage{graphicx}
\usepackage{dirtytalk}
\usepackage{csquotes}
\usepackage{amsfonts}
\usepackage[utf8]{inputenc}
\usepackage{graphicx}
\usepackage{eucal}
\usepackage{amsmath}
\usepackage{amsfonts}
\usepackage{amssymb}
\usepackage{mathtools}
\usepackage{tabu}
\usepackage{tabulary}
\usepackage{booktabs}
\usepackage{caption}
\usepackage{subcaption}
\usepackage{empheq}
\usepackage{eqparbox} 
\usepackage{framed}
\usepackage{algpseudocode}

\usepackage{xurl}
\usepackage{wrapfig}
\usepackage{lscape}
\usepackage{rotating}
\usepackage{epstopdf}
\usepackage{pgfgantt}
\usepackage{ragged2e}
\usepackage{geometry} 
\usepackage{MnSymbol}
\usepackage{pdflscape}
\usepackage{tcolorbox}
\usepackage{svg}
\usepackage{subcaption}

\newganttchartelement*{mymilestone}{
mymilestone/.style={
shape=isosceles triangle,
inner sep=0pt,
draw=cyan,
top color=white,
bottom color=cyan!50
},
mymilestone incomplete/.style={
/pgfgantt/mymilestone,
draw=yellow,
bottom color=yellow!50
},
mymilestone label font=\slshape,
mymilestone left shift=0pt,
mymilestone right shift=0pt
}

\newgantttimeslotformat{stardate}{%
\def\decomposestardate##1.##2\relax{%
\def\stardateyear{##1}\def\stardateday{##2}%
}%
\decomposestardate#1\relax%
\pgfcalendardatetojulian{\stardateyear-01-01}{#2}%
\advance#2 by-1\relax%
\advance#2 by\stardateday\relax%
}

\usepackage{pgfplots}
\pgfplotsset{compat=1.15}

\makeindex						

\begin{document}

	\author{YOUSEF EMAMI}
	\title{Deep Reinforcement Learning for Joint Cruise Control and Intelligent Data Acquisition in UAVs-Assisted Sensor Networks}

	\thesisdate{December 2023}


	\begin{Prolog}
	\thispagestyle{empty}

\chapter*{Abstract}

Unmanned aerial vehicle (UAV)-assisted sensor networks (UASNets), which play a crucial role in creating new opportunities, are experiencing significant growth in civil applications worldwide. UASNets provide a range of new functionalities for civilian sectors. Just as UASNets have revolutionized military operations with improved surveillance, precise targeting, and enhanced communication systems, they are now driving transformative change in numerous civilian sectors. For instance, UASNets improve disaster management through timely surveillance and advance precision agriculture with detailed crop monitoring, thereby significantly transforming the commercial economy. UASNets revolutionize the commercial sector by offering greater efficiency, safety, and cost-effectiveness, highlighting their transformative impact. A fundamental aspect of these new capabilities and changes is the collection of data from rugged and remote areas. Due to their excellent mobility and maneuverability, UAVs are employed to collect data from ground sensors in harsh environments, such as natural disaster monitoring, border surveillance, and emergency response monitoring. One major challenge in these scenarios is that the movements of UAVs affect channel conditions and result in packet loss. Fast movements of UAVs lead to poor channel conditions and rapid signal degradation, resulting in packet loss. On the other hand, slow mobility of a UAV can cause buffer overflows of the ground sensors, as newly arrived data is not promptly collected by the UAV.

Our proposal to address this challenge is to minimize packet loss by jointly optimizing the velocity controls and data collection schedules of multiple UAVs. The states of ground sensors include battery level, data queue length, and channel quality. In the absence of up-to-date knowledge of ground sensors’ states, we propose a multi-UAV deep reinforcement learning-based scheduling algorithm (MADRL-SA). This algorithm allows UAVs to asymptotically minimize packet loss due to buffer overflows and poor channel conditions, even in the presence of outdated knowledge of the network states at individual UAVs.

Furthermore, in UASNets, swift movements of UAVs result in poor channel conditions and fast signal attenuation, leading to an extended age of information (AoI). In contrast, slow movements of UAVs prolong flight time, thereby extending the AoI of ground sensors. Additionally, the UAVs should consider the movements of other UAVs to minimize the average AoI by coordinating their velocities. Hence, finding an equilibrium solution among UAVs to optimize velocity and reduce the average AoI becomes crucial.

To address this challenge, we propose a new mean-field flight resource allocation optimization to minimize the AoI of sensory data. Balancing the trade-off between UAV movements and AoI is formulated as a mean-field game (MFG). We introduce a new mean-field hybrid proximal policy optimization (MF-HPPO) scheme to handle the expanded solution space of MFG optimization. This scheme minimizes the average AoI by optimizing the UAV trajectories and ground sensor data collection schedules, considering mixed continuous and discrete actions. Additionally, we incorporate a long short-term memory (LSTM) in MF-HPPO to predict the time-varying network state and stabilize the training.

\vspace*{10mm}

\vspace*{15mm}

\textbf{Keywords: UAVs, Mean-field game, Age of information, Proximal policy optimization, Long short term memory, Communication scheduling, Velocity control, Deep Q-Network.}  						
		\clearemptydoublepage
		
		\input{} 							
		
		\input{}  						
		
		\input{}  					
		
		
		\pdfbookmark[0]{Table of Contents}{contents} 			
		\tableofcontents
		\clearemptydoublepage

		\listoffigures
		\addcontentsline{toc}{chapter}{\listfigurename}			
		\clearemptydoublepage
		
		\listoftables
		\addcontentsline{toc}{chapter}{\listtablename}			
		\clearemptydoublepage
		
		\thispagestyle{empty}

\chapter*{List of Abbreviations}
\chaptermark{List of Abbreviations}

\begin{flushleft}
	\begin{longtable}{l p{0.8\linewidth}}
		UAV & \hspace{5mm} Unmanned Aerial Vehicle \\[1mm]
        DRL & \hspace{5mm} Deep Reinforcement Learning \\[1mm]
		DQN	& \hspace{5mm} Deep Q-Network\\[1mm]
		DDPG	& \hspace{5mm} Deep Deterministic Policy Gradient \\[1mm]
		PPO & \hspace{5mm} Proximal Policy Optimization\\[1mm]
		MFG & \hspace{5mm} Mean Field Game \\[1mm]
        RL & \hspace{5mm} Reinforcement Learning \\[1mm]
        MDP& \hspace{5mm} Markov Decision Process \\[1mm]
        
		MMDP & \hspace{5mm} Multi-Agent Markov Decision Process \\[1mm]
        AoI & \hspace{5mm} Age of Information\\[1mm]
		LSTM	& \hspace{5mm} Long Short Term Memory\\[1mm]
		UASNets & \hspace{5mm} UAVs-Assisted Sensor Networks  \\[1mm]
		FPK	& \hspace{5mm} Fokker-Planck-Kolmogorov  \\[1mm]
		LoS & \hspace{5mm} Line-of-Sight\\[1mm]
		QoS & \hspace{5mm} Quality of Service \\[1mm]
		UMi & \hspace{5mm} Urban Micro \\[1mm]
        UMa & \hspace{5mm} Urban Macro \\[1mm]
		SWAP & \hspace{5mm} Size, Weight, and Power \\[1mm]
        TRPO& \hspace{5mm} Trusted Region Policy Optimization \\[1mm]
		MARL & \hspace{5mm} Multi-Agent Reinforcement Learning \\[1mm]
		Dec-POMDP & \hspace{5mm} Decentralized Partially Observable Markov Decision Process \\[1mm]
		MADRL-SA & \hspace{5mm} Multi-UAV Deep Reinforcement Learning based Scheduling Algorithm \\[1mm]
		MF-HPPO & \hspace{5mm} Mean Field Hybrid Proximal Policy Optimization\\[1mm]
		WSNs & \hspace{5mm} Wireless Sensor Networks.\\[1mm]
		ABS & \hspace{5mm} Aerial Base Station\\[1mm]
        DoF & \hspace{5mm} Degree of Freedom \\[1mm]
        ML & \hspace{5mm} Machine Learning \\[1mm]
	    IEEE & \hspace{5mm} Institute of Electrical and Electronics Engineers \\[1mm]
	    RNN & \hspace{5mm} Recurrent Neural Network \\[1mm]
        6G & \hspace{5mm} 6 Generation \\[1mm]
        5G & \hspace{5mm} 5 Generation \\[1mm]
        eMBB & \hspace{5mm} Enhanced Mobile Broadband \\[1mm]
	 URLLC & \hspace{5mm} Ultra Reliable Low Latency Communications \\[1mm]
      mMTC & \hspace{5mm} Massive Machine Type Communications\\[1mm]
       EU & \hspace{5mm} European Union \\[1mm]
      IoT & \hspace{5mm} Internet of Things \\[1mm]
      UAS & \hspace{5mm} Unmanned Aerial Systems \\[1mm]
      US & \hspace{5mm} United States \\[1mm]
      DNN & \hspace{5mm} Deep Neural Network \\[1mm]
		\end{longtable}
\end{flushleft}
		\addcontentsline{toc}{chapter}{List of Abbreviations}
		\clearemptydoublepage	
	\end{Prolog}

	\StartBody
         \chapter{Introduction}
\label{chap:Intro}
Unmanned aerial vehicles (UAVs) have become indispensable in today’s technological advancements, bringing about significant changes in various fields. They have revolutionized sectors such as agriculture,\cite{kurunathan2023machine},\cite{li2017wireless}, public safety\cite{li2021continuous},\cite{li2018reinforcement}, environmental monitoring\cite{li2020deep},\cite{li2022deep}, and security\cite{10352334}, \cite{wang2017pele},\cite{wang2019eavesdropping}. In the realm of agriculture, UAVs hold great potential for precision farming, aligning with the European Union’s focus on sustainable and environmentally friendly agricultural practices. Additionally, UAVs have proven their worth in assessing hazardous situations\cite{noor2021hybrid}, conducting search and rescue missions\cite{li2020multi}, gathering evidence for investigations \cite{li2021practical}, and detecting potential threats \cite{li2019proactive},\cite{li2021bloothair},\cite{undertaking2017european}. Furthermore, UAVs play a crucial role in the development of 5th generation (5G) networks, contributing to the realization of 5G’s goals, including enhanced mobile broadband (eMBB), ultra-reliable and low latency communications (URLLC), and massive machine-type communications (mMTC). In the context of eMBB, UAVs provide high data rates, particularly in densely populated or remote areas. They can act as aerial base stations (ABS) or relays, supporting URLLC and reducing latency for real-time communication. Moreover, UAVs facilitate mMTC by enabling the deployment of Internet of Things (IoT) devices in challenging environments and optimizing network resources to handle a large number of connections. Looking ahead, UAVs are expected to play a pivotal role in 6th generation (6G) networks, enabling improved data collection and analysis. Enhanced data collection techniques allow for real-time capture of a wider range of data, thereby enhancing decision-making processes. This opens up opportunities for precise environmental monitoring, real-time traffic analysis, and prompt disaster response through immediate aerial assessments  \cite{li2018uav}, \cite{JIANG202219}.

UAVs have the capability to operate in challenging and remote environments, making them ideal for aerial data collection. The integration of UAVs into sensor networks for this purpose is known as UAVs-assisted sensor networks (UASNets). UAVs can serve as aerial base stations (ABS) or relays to extend the coverage and connectivity of sensor networks \cite{mozaffari2019tutorial},\cite{li2020onboard}. The advancement in UAV manufacturing and the miniaturization of communications equipment have made it possible to incorporate compact and lightweight wireless transceivers into UAVs, enabling efficient aerial data collection. Commercial wireless transceivers suitable for UAV installation with moderate payloads are already available in the market\cite{li2020onboard2},\cite{7249319}. Compared to traditional terrestrial communications that rely on fixed gateway locations, UASNets offer several advantages. Firstly, aerial data collectors can be rapidly deployed, making them particularly beneficial for harsh and remote areas. Secondly, due to their high altitude, UAVs have a higher probability of establishing line-of-sight (LoS) connections with ground sensors, resulting in more reliable communication links. Thirdly, the mobility of UAVs provides an additional degree of freedom (DoF) for optimizing communication performance by dynamically adjusting their positions in three dimensions to meet the communication demands on the ground.

Integrating UAVs into wireless sensor networks (WSNs) presents new design opportunities but also brings challenges. UASNets differ significantly from terrestrial networks due to factors such as the high altitude and mobility of UAVs, the likelihood of LoS channels between UAVs and ground sensors, varying quality of service (QoS) requirements for payload and non-payload data, strict size, weight, and power (SWAP) constraints of UAVs, and the need to jointly optimize UAV mobility control and communication scheduling/resource allocation to maximize system performance.
\begin{itemize}
    \item High altitude: UAV data collectors are positioned at much higher altitudes compared to traditional terrestrial gateways. While terrestrial gateways are typically located at around 10m for urban micro deployment and 25m for urban macro deployment, UAVs can fly as high as 122m under current regulations. This higher altitude enables UAV data collectors in UASNets to achieve wider ground coverage compared to their terrestrial counterparts.
    
    \item Higher channel gain: The air-ground channels experienced by UAVs exhibit distinct characteristics due to their high altitude. Unlike terrestrial channels that suffer from low channel gain due to shadowing and multipath fading, UAV ground sensor channels generally have limited scattering and primarily rely on LoS links, resulting in higher channel gain. This LoS-dominant air-ground channel offers more reliable link performance between UAVs and associated ground sensors.

    \item Controlled mobility: Unlike fixed terrestrial gateways, UAVs possess the capability to move at high speeds in three-dimensional space, allowing for controlled mobility. While this mobility introduces time-varying channels with ground sensors, it also opens up new design opportunities for communication-aware control of UAV mobility. UAVs can optimize their position, altitude, velocity, heading direction, and trajectories to adapt to communication objectives and improve overall network performance.
   
    \item SWAP constraints: UAVs face significant SWAP constraints, which limit their endurance, computational capacity, and communication capabilities. Unlike terrestrial communications systems that benefit from stable power supplies at fixed gateways, UAVs must operate within these constraints, requiring efficient power management, lightweight hardware, and optimized communication protocols \cite{ruizhang},\cite{li2020deep2}.
    \end{itemize}

Meanwhile, UASNets enhance the decision-making process through their advanced data collection capabilities. By gathering comprehensive and real-time information, they provide a rich and accurate basis for decision-making. These networks combine the agility and adaptability of UAVs with the extensive data collection capabilities of ground sensors, creating a system that not only collects valuable data but also reacts quickly and adjusts to changing environmental conditions. This adaptability makes UASNets highly effective in dealing with different situations\cite{kurunathan2021deep},\cite{li2022data2}. The following reasons highlight the importance of UASNets as a significant research area: (i) UASNets can cover large areas and collect high-quality data in real time. This makes them valuable in various fields such as environmental monitoring, wildlife protection, and infrastructure inspection. (ii) UASNets play a critical role in providing vital information to first responders during natural disasters. This information enriches the decision-making process and allows for more efficient resource allocation. (iii) Farmers can efficiently monitor crop health, soil conditions, and water needs using UASNets. This improves agricultural quality, increases productivity, and reduces environmental impact. (iv) UASNets enable agencies to regularly inspect critical infrastructure such as bridges, dams, and power lines. This proactive approach reduces the risk of catastrophic failures. In summary, research on UASNets contributes to the development of innovative solutions for practical problems, improves quality of life, and promotes sustainable development.
The EU recognizes the importance of UASNets and their integration into various disciplines. The following are reasons highlighting the importance of UASNets in the EU: The EU is committed to sustainable development and environmental preservation. UASNets play a crucial role in providing important data for monitoring air pollutants, identifying their sources, and assessing ecosystem health. This aligns with the EU’s goals of reducing emissions and preserving the environment. The EU has an aging infrastructure network that requires regular inspection and maintenance. UASNets can help take proactive measures by identifying potential problems, enabling timely maintenance, and reducing the risk of catastrophic failures. This contributes to ensuring public safety. With a projected fleet of approximately 50,000 UAVs, UASNets can support public safety missions. UASNets can be utilized in the energy sector for performing preventative maintenance inspections and mitigating risks to personnel and infrastructure. It is estimated that around 10,000 UAVs will be used in this sector, contributing to efficient and safe energy operations. The EU envisions a fleet of 400,000 UAVs for civil applications by 2050. By leveraging UASNets, the EU can address various challenges while promoting sustainability, economic growth, and improved quality of life for its citizens.

 Europe is not the only region making intensive efforts to utilize UAVs. The United States (US) and China, two major countries, are investing significantly in technology and innovative companies, surpassing the level of European investments. Specifically, the US and China are leaders in the production of defense and civil UAVs  \cite{undertaking2017european}. This emphasizes the transformative potential of UASNets in addressing technological complexities and economic constraints.

A practical example of UASNets can be observed in precision agriculture. The role of agriculture is of paramount importance to the European economy, with food security being a top priority. UASNets can optimize agricultural practices, minimize waste, and enhance crop productivity, thereby contributing to the overall goals of the European agricultural sector. It is predicted that in the agriculture sector, more than 100,000 UAVs will enable precision agriculture to achieve the necessary increase in productivity.

Fig. \ref{fig:application} shows a typical UASNets setup where ground sensors monitor farmland. The ground sensors generate sensory data, which is stored in a data queue for later transmission to the UAVs. The UAVs hover over the farmland, approaching each ground sensor closely to collect data over short distances. In this scenario, a large farm is equipped with soil sensors that continuously monitor various parameters, including soil moisture, temperature, and nutrient levels. These ground sensors consistently gather data, providing farmers with information about irrigation, fertilization, and crop protection

UAVs are employed as aerial data collectors, patrolling over the farmland and utilizing LoS communications. They manage their mobility to approach ground sensors and collect sensory data. UAVs can improve overall network coverage and performance, enabling farmers to access comprehensive and accurate data for optimizing their farming practices.

The remainder of this chapter is organized as follows: Section \ref{chap1:sec2} presents the motivation for this work. Section \ref{chap1:sec3} outlines the thesis and research questions. Section \ref{chap1:sec4} presents the methodology. Section \ref{chap1:sec5} outlines the structure of the thesis.
\section{Motivation}  \label{chap1:sec2}
Thanks to their exceptional mobility and maneuverability, UAVs are utilized in various civil and commercial applications, including weather monitoring, traffic control, package delivery \cite{guizani},\cite{li2023exploring}, and crop monitoring \cite{crop}. They are also employed as data relays for ground sensors in challenging environments such as natural disaster monitoring \cite{disaster},\cite{ma2022robust}, border surveillance \cite{civil},\cite{li2019energy}, and emergency assistance \cite{emergency},\cite{guan2023mappo}. In scenarios where ground sensors are deployed beyond the reach of terrestrial gateways and lack a consistent power supply, UAVs can physically approach each individual ground sensor. The short-range LoS communication link between a UAV and a ground sensor exhibits significant channel gain, enabling high-speed data transmission. By utilizing UAVs for data collection, network throughput can be improved, and coverage can be extended beyond terrestrial gateways. Moreover, UASNets offer several advantages for data collection in remote and inhospitable environments. UAVs can access areas that are challenging for humans to reach, resulting in more efficient and cost-effective data collection. This approach reduces safety risks as the use of UAVs eliminates the need for human intervention in hazardous environments. Due to their mobility, UAVs have the capability to cover vast areas, thereby reducing the time and resources required for data collection.

\begin{figure*} 
        \centering
    	\includegraphics[ width=5.5in, height =3in]{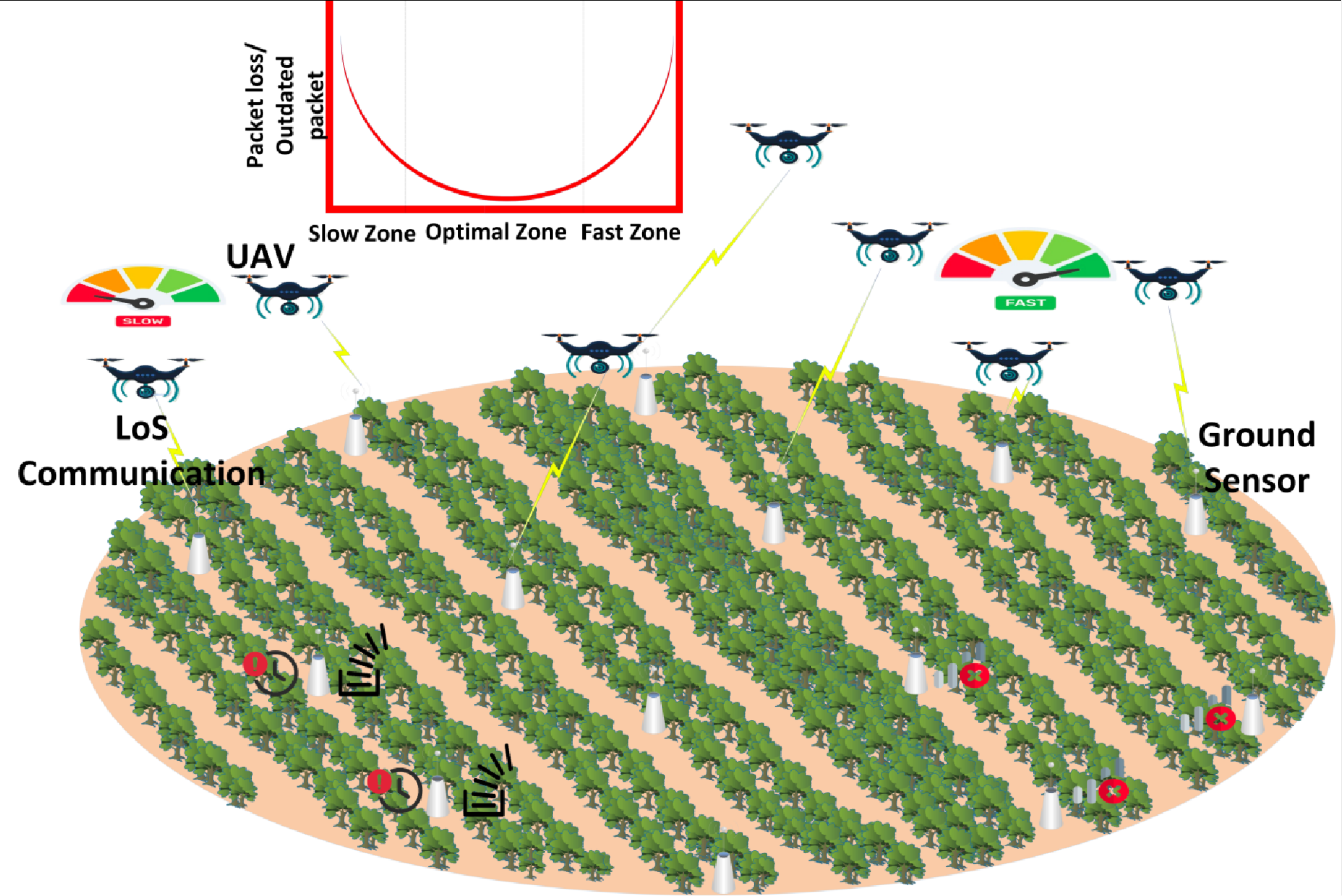}
    	\caption{An overview of UASNets for precision agriculture.}
    	\label{fig:application}
\end{figure*}
In UASNets, ground sensors are exposed to random data inputs as data generation is influenced by unpredictable variations in temperature and humidity. As depicted in  Fig. \ref{fig:application}, UAVs are deployed to hover over farmland, allowing close proximity to ground sensors and utilizing short LoS communication links for data collection. However, selecting a ground sensor for data collection may lead to buffer overflows for other sensors if their buffers are already full while new data continues to arrive. Moreover, transmissions from ground sensors located far away from the UAVs, experiencing poor channel conditions, are susceptible to errors at the UAVs. The slow mobility of a UAV can contribute to buffer overflows in ground sensors as newly arrived data is not promptly collected by the UAV. Properly scheduling data collection, taking into account the onboard velocity of the UAVs, is crucial to avoid data queue overflow and communication failures. Additionally, coordination between participating UAVs is necessary for joint velocity control and sensor selection. However, real-time sharing of velocities and selected sensors among UAVs is challenging due to limited radio coverage and the fast movements of UAVs.

In summary, the effective management of joint communication scheduling and velocity control is crucial to minimize packet loss, preventing buffer overflows and communication failures in UASNets. However, it is important to note that ensuring the freshness and relevance of collected data is also essential. To achieve this, minimizing the age of information (AoI) becomes necessary in UASNets.

In UASNets, the AoI is commonly used to measure the freshness of sensory data \cite{kaul2012real} collected at ground sensors and received by the UAVs. It represents the time elapsed between data generation at a ground sensor and its receipt at the UAV, accounting for transmission time and network delays. When the UAV’s flight is not properly controlled, it can move away from the ground sensor, increasing the AoI and causing data to expire. Additionally, different ground sensors may have varying AoI due to the impact of monitored natural conditions on data generation  \cite{aggregator}. The optimization of UAV cruise control and communication schedules to minimize AoI becomes challenging because the UAV has limited knowledge of ground sensors’ data generation rate and channel conditions. Swift movements of the UAV result in poor channel conditions and frequent data retransmissions, leading to a prolonged AoI. Conversely, slow UAV movements extend flight time and increase the AoI of ground sensors. Furthermore, the UAV needs to consider the movements of other UAVs to minimize the average AoI by coordinating their velocities, highlighting the importance of finding an equilibrium solution.

Decentralized approaches are relevant when UAVs have limited information about each other’s actions, such as trajectory, flight speed, and scheduled ground sensors. Game theory can be applied to design decentralized control and determine equilibria in UAV networks \cite{8723552}. However, traditional game theory approaches become computationally intractable with a large number of UAVs. Mean-field game (MFG), on the other hand, offers a scalable framework to address the joint optimization of cruise control and communication schedules. MFG approximates the interactive behavior of a large number of UAVs using a continuum or mean field, significantly reducing computational complexity. It enables UAVs to make decisions based on the overall swarm behavior rather than individual actions.

\section{Thesis Statement and Research Questions} \label{chap1:sec3}
\label{sec:introduction:thesis_statement}
In this research, our proposed solutions aim to address the challenges faced by UASNets. In this framework, the thesis statement is as follows:

\vspace{0.2cm}
\noindent \fbox{\begin{tcolorbox}
\textit{We postulate that incorporating cruise control and data collection scheduling into UASNets can effectively alleviate the impact of channel conditions and unlock the advantages of timeliness and resource utilization in UASNets.}
\end{tcolorbox}}
\vspace{0.2cm}

Based on this thesis, our research focuses on investigating and enhancing the performance of data collection in UASNets. We envision that by adopting this new paradigm, real-time decision-making can be facilitated, leading to improved resource utilization. Consequently, we anticipate advancements in QoS, overall system reliability, and productivity of UASNets. However, achieving this goal entails addressing several challenging scientific problems, which we formulate as the following two research questions.
\begin{enumerate}
    \item[$(\operatorname{RQ}_1)$.] \textbf{Research Question~1:}  How can we develop a joint communication scheduling and velocity control mechanism for data collection in UASNets to minimize packet loss and mitigate the impact of UAV movement on data transmission, while addressing the challenges posed by limited radio coverage and rapid movement?

\vspace{0.2cm}
\begin{tcolorbox}
    How the joint communication scheduling and velocity control mitigate the effects of UAVs’ movement on data transmission in the presence of fast movements.
\end{tcolorbox} 
\vspace{0.2cm}

    \item[$(\operatorname{RQ}_2)$.] \textbf{Research Question~2:} In the presence of a large number of UAVs, the challenge lies in developing cruise control mechanisms that minimize the AoI and mitigate the impact of UAVs’ movements on AoI. Additionally, how can we find an equilibrium solution and capture the temporal dependencies of cruise control?

\vspace{0.2cm}
\begin{tcolorbox}
       How can we develop cruise control and mitigate the impact of UAV move ments on AoI.
\end{tcolorbox}
\vspace{0.2cm}

\end{enumerate}
\section{Methodology}   \label{chap1:sec4}
This work focuses on improving the performance of data collection in UASNets, particularly in lossy channels. The main objective is to minimize packet loss and AoI in order to enhance the efficiency of data collection. To achieve this goal, the study explores the application of deep reinforcement learning (DRL). By leveraging DRL algorithms, the research aims to develop intelligent and adaptive solutions that can learn from the environment and determine the optimal policy.
 
The use of DRL-based techniques is expected to provide valuable insights and effective approaches to significantly enhance the performance of data collection in UASNets. The ultimate aim is to contribute to the advancement of this emerging field by proposing novel solutions that leverage DRL for improved data collection performance.

In this thesis, the joint communication schedule and velocity control of multiple UAVs are formulated as a 	multi-agent Markov decision process (MMDP) to minimize packet loss caused by buffer overflows and communication failures. The ground sensor keeps a record of the visit time whenever a UAV schedules data transmission from the sensor. Furthermore, the visiting records of the sensor are shared with the UAV, serving as evidence of other UAVs’ communication schedules. The network state in the MMDP includes battery levels and data queue lengths of the ground sensors, channel conditions, visit time, and waypoints along the UAVs’ trajectories. The UAVs take actions such as selecting ground sensors for data transmission, determining modulation schemes, and adjusting patrol velocities. In practical scenarios, the UAVs lack real-time knowledge of the battery level and data queue length of the ground sensors. Thus, multi-UAV Q-learning can be employed to train the UAVs’ actions. However, since each UAV’s trajectory may have a large number of waypoints, controlling the velocities of the UAVs along the trajectories results in a vast state and action space, making multi-UAV Q-learning complex.

In our MFG approach, the optimal velocities of the UAVs are determined by solving a Fokker–Planck–Kolmogorov (FPK) equation. This equation describes the evolution of the mean field to achieve an equilibrium of the optimal velocities of the UAVs. However, in practical scenarios, the proposed MFG solution is challenging to implement online due to the lack of instantaneous knowledge of the UAV’s cruise control decisions and AoI. To address this, we formulate the flight resource allocation optimization problem in the MFG framework as an MMDP. The network states in the MMDP consist of the AoI of the ground sensors and the waypoints of the UAV swarm. The action space in the MMDP includes continuous variables such as waypoints and velocities, as well as discrete variables representing transmission schedules. To tackle this complex problem, we propose a solution called the mean-field hybrid proximal policy optimization (MF-HPPO). MF-HPPO aims to optimize both the cruise control of the UAVs and the transmission schedules of the ground sensors in a coordinated manner, leveraging the advantages of the mean-field approximation.

The research topics of this thesis can be summarized as follows:
\begin{itemize}
    \item Joint velocity control and communication scheduling to minimize packet loss and using DRL to find the optimal policy.
    \item Cruise control based on MFG to minimize AoI and using DRL to find the mean field equilibrium.
    
\end{itemize}

\section{Contributions}  \label{chap1:sec5}

In this section, we summarize the main findings of our research in relation to the research questions outlined in Section \ref{chap1:sec3}  and discuss the contribution of our work to the existing body of knowledge in the field of data collection.

\begin{enumerate}
    \item This contribution addresses RQ1. The problem of joint velocity control and data collection scheduling is formulated as an MMDP to minimize packet loss caused by buffer overflow and channel fading. To handle the large state and action spaces, we propose the multi-UAV DRL-based scheduling algorithm (MADRL-SA) using Deep-Q-Networks (DQN) to optimize the selection of ground sensors, instantaneous patrol velocities of UAVs, and modulation schemes. The inclusion of experience replay enhances the learning efficiency of the algorithm by reducing sample correlations.
    
    The mentioned contribution is of utmost importance as it addresses the challenges faced by modern UAV networks in handling complex dynamic environments, including UAV movement. The proposed methodology showcases the potential of DRL in solving complex problems. It also contributes to the development of intelligent, adaptive, and autonomous systems capable of self-optimization. The use of DRL in conjunction with experience replay enhances the system’s ability to learn and evolve, leading to improved performance. 

\item This contribution addresses RQ1. : In practice, the online decisions of UAVs during flight are unknown to each other, which can hinder the training of MADRL-SA. To address this, a local action recording process is developed where ground sensors record historical visits of all UAVs. The UAV scheduling a ground sensor receives these records, providing information about the past scheduling decisions of other UAVs.

The introduction of the local action recording process in this contribution is an important step in addressing the practical challenges associated with UASNets. In practical scenarios, UAVs are unaware of each other’s decisions, leading to uncertainty and potentially incomplete training of MADRL-SA. This situation can result in suboptimal decisions and degrade network performance. By incorporating a local action recording process, the algorithm’s effectiveness is ensured even under realistic operating conditions. This approach promotes a more collaborative environment among UAVs, allowing them to adjust their actions based on the observed behavior of other UAVs in the network. 

\item This contribution addresses RQ2. A novel formulation of MFG optimization with a large number of UAVs is proposed to address the trade-off between UAV cruise control and AoI. Due to the computational complexity of MFG, the MF-HPPO algorithm is introduced to minimize average AoI. The algorithm learns state dynamics and optimizes UAV actions in a mixed discrete and continuous action space.

This contribution represents a significant advancement in the field of UASNets with a large number of UAVs. By leveraging MFG optimization, we effectively address the challenges associated with managing such complex systems. Our approach focuses on the collective behavior of UAVs, leading to improved resource allocation and overall performance. MF-HPPO efficiently optimizes both continuous and discrete actions to minimize the average AoI. This ability to optimize UAV actions in a mixed-action space highlights their versatility and adaptability, enabling them to meet diverse network conditions and requirements. The proposed method underscores the importance of advanced optimization techniques in solving complex real-world problems. Moreover, this contribution pushes the boundaries of UASNets and highlights the wider applicability of MFG optimization in addressing complex challenges across various domains.
\item 
This contribution addresses RQ2. To capture temporal dependencies in cruise control and improve learning convergence, a new long short-term memory (LSTM) layer is developed within the MF-HPPO algorithm. This LSTM layer predicts time-varying network states, such as AoI and UAV waypoints.

By incorporating the LSTM layer, our contribution tackles the issue of capturing temporal dependencies in cruise control, which is crucial for efficiently managing and optimizing UASNets. This development emphasizes the significance of combining advanced machine learning (ML) techniques to create 
more robust and adaptable algorithms.
\end{enumerate}

\section{Thesis Structure} \label{chap1:sec6}
The rest of this document is structured as follows.
\begin{itemize}
    \item Chapter 2 discusses background and related work. In this chapter, we present background on DRL  and then delve into the existing literature to explore related work on flight resource allocation and scheduling. Our objective is to gain a comprehensive understanding of the relevant research in order to comprehend the joint communication scheduling and velocity control in UASNets. Additionally, we review the literature on mean-field flight resource allocation and time-critical flight resource allocation to analyze their strengths and weaknesses. This analysis serves as a foundation for developing a cruise control system based on MFG theory that minimizes AoI.
    \item Chapter 3 formulates the joint communication scheduling and velocity control problem as an MMDP to minimize packet loss resulting from communication failures and buffer overflows. Given the large state and action spaces, we employ DRL techniques to discover the optimal policy for this problem.
    \item Chapter 4 formulates cruise control based on MFG theory to minimize AoI. Solving the MFG problem online poses challenges, hence we formulate it as an MMDP, encompassing both continuous and discrete actions. To address this MMDP formulation, we propose the MF-HPPO algorithm, which optimizes actions in a mixed-action space. 
\end{itemize}							%
	   \clearemptydoublepage
	   \chapter{Background and Related work} \label{related work}

UASNets have emerged as an innovative technology that offers enhanced data collection capabilities for various applications, including environmental monitoring, disaster management, and surveillance. In UASNets, UAVs play a critical role in gathering sensory data. However, a significant challenge in UASNets is the dynamic nature of UAV movements, which greatly impacts channel conditions and gives rise to issues such as packet loss and outdated packets. The rapid movements of UAVs can result in unfavorable channel conditions and quick signal degradation, requiring frequent data retransmissions. Conversely, slow movements prolong the flight time, leading to delays in collecting newly arrived data by the UAV. To tackle these challenges, one potential strategy is to perform joint cruise control and communication scheduling in the presence of lossy channels, aiming to minimize packet loss and AoI.
One effective approach for addressing the challenges in UASNets is cruise control and data collection scheduling. In the following section, we present background information on DRL then we provide a review of the existing literature on this problem.
The relevant state-of-the-art works in this area can be classified into three categories: i) DRL-aided flight resource allocation and scheduling, ii) DRL-aided flight resource allocation using MFG, and iii) DRL-aided flight resource allocation for data freshness. 
\section{Deep Reinforcement Learning}

DRL is a prominent branch of ML in which an agent learns to interact within an environment by taking actions and observing the resulting outcomes \cite{sutton2018reinforcement}.

DRL is particularly useful for solving MMDPs that have unknown transition probabilities. During the DRL process, an agent observes its current state, selects an action, and receives immediate feedback in the form of a cost or reward, along with the new state. The observed information, such as the immediate cost and new state, is then utilized to adjust the agent’s policy. This iterative process continues until the agent’s policy converges toward the optimal policy \cite{asurvey}.
DRL can be applied to UASNets for the following reasons: (a) UAVs may face challenges in implementing mathematical models of the complex environment or may not have access to such models. (b) The mobility feature of UAVs leads to large state spaces and action spaces. (c) UAVs often lack up-to-date knowledge about the status of ground sensors, including battery, energy, and channel conditions.

Formally, DRL can be described as an MMDP, which includes the number of agents, state, action, shared cost function, and transition probability. An MMDP is a mathematical framework used to model decision-making in situations where multiple agents interact with each other in an uncertain environment. In an MMDP, the action taken by each agent not only determines the future state but also affects the actions of other agents. Furthermore, in an MMDP, a shared cost function is employed. This shared cost function considers the joint action of agents and provides feedback that is common to all agents. The objective is to encourage collaboration among the agents toward a shared goal, rather than individual goals that may conflict with each other. Well-designed shared cost functions can foster collaboration among agents and lead to more favorable outcomes for the entire team. The MMDP framework finds applications in various domains, including UAV swarm control and multiplayer games..

Q-learning, due to the exponential growth of states and actions caused by the mobility of UAVs, is unable to handle the resource allocation problem in UASNets. This issue, commonly referred to as the curse of dimensionality, poses a significant challenge. However, DQN offers a solution to overcome this challenge. In the context of UASNets, DQNs play a crucial role in optimizing various operational aspects, including flight trajectory, cruise control, and data collection scheduling. They employ deep neural networks to represent the action-value function of each agent. The state information captured by the DQN can encompass the UAV’s current location, the status of sensor nodes, and the traffic conditions within the network. The available actions can involve adjusting parameters such as speed, trajectory, or data transmission schedules. DQN incorporates the use of a target network and experience replay for each UAV to ensure stability in the learning process. Experience replay is utilized in DQN to eliminate correlations in the observation sequence and mitigate abrupt changes in the data distribution by randomizing states and actions within the MMDP. Consequently, DQN aids in forming a policy that minimizes the cost function and enhances the overall performance of UASNets.

DQN is primarily designed to optimize discrete actions and is limited in its ability to handle continuous actions. To address this limitation and optimize both discrete and continuous actions, proximal policy optimization (PPO) can be employed. In the context of UASNets, PPO plays a crucial role in enhancing the operational efficiency of these networks. As a type of policy gradient method, PPO enables the optimization of UAV trajectories (continuous action space) and data collection schedules (discrete action space). The algorithm maintains a delicate balance between exploring new strategies and exploiting the current strategy, which is particularly advantageous in complex and dynamic environments like UASNets.

PPO achieves stability and efficient learning by ensuring only a small deviation from the previous strategy during each update, thereby mitigating the risk of detrimental updates. The objective of PPO is to minimize costs, which can be tailored to reflect real-time data collection requirements. By optimizing both trajectories and data collection schedules, PPO facilitates the development of robust policies that enhance the overall performance of UASNets. PPO has two primary variants: PPO-penalty modifies the hard constraint of TRPO by incorporating it as a penalty in the objective function. On the other hand, PPO-clip does not impose a constraint but utilizes clipping techniques to bind the changes in the policy.
\section{DRL-aided Flight resource allocation and scheduling} \label{chap2:sec1}
The work in \cite{zheng} develops a framework for trajectory control, user association, and power control in  multi-UAV enabled wireless networks. Communication throughput gains can be obtained by mobile UAVs over static UAVs/fixed terrestrial base stations, by exploiting the design DoF via UAV trajectory adjustment. A general mixed integer nonlinear program formulation for a multi-UAV network is presented in \cite{computation} to adjust the communication and the computational energy.  \cite{jointposition} explores a multi-UAV-aided relaying system, where  UAV relays aim to establish communication between senders and receivers and  to improve the rate between the pair of sender and receiver, the UAVs’ positions are adjusted, and resource allocations are conducted.
In \cite{sharma2016energy}, a cooperative framework designed which allowed the formation of a network between the aerial and the ground nodes. Their approach provides continuous connectivity, enhanced lifetime, and improved coverage in the UAV coordinated WSNs and laid the foundation of guided network formations between the UAVs and the ad hoc networks on the ground. A  framework  is developed in \cite{deadline} to improve energy efficiency in deadline-based WSN data collection with multiple UAVs. In \cite{completion}, the mission completion time is adjusted for multi-UAV-enabled data collection. An  energy-efficient  transmission scheduling scheme of UAVs in a cooperative relaying network  is developed in \cite{kairelay} such that the maximum energy consumption of all the UAVs is  minimized,  in  which  an  applicable  sub-optimal  solution is developed  and  the  energy  could  be  saved  up  to  50\% via  simulations. In \cite{targetdeadline} a UAV is used to collect data from time-constrained Internet of Things (IoT) devices. The UAV trajectory and radio resource allocation are adjusted to collect data from IoT devices, adapting to their deadline.  

In\cite{dqn},  a single-agent  DQN for UAV-assisted online power transfer and data collection is developed. However, in most situations, multiple UAVs are needed to interact with each other to solve a resource allocation problem. In \cite{obdrl}, online velocity control and data capture are studied in UAV-enabled IoT networks. DQN is developed in the presence of outdated knowledge to determine the patrolling velocity and data transmission schedule of the IoT node. In \cite{kaicollection}, the joint flight cruise control and data collection scheduling in the UAV-aided IoT network is formulated as a POMDP to minimize the data lost due to buffer overflows at the IoT nodes and fading airborne channels. A UAV-assisted IoT communication is investigated in \cite{iotconf}  where by applying multi-agent DRL a resource allocation scheme adapting to bandwidth, throughput, and interference is obtained. A wireless powered communication network is developed in \cite{wpcn} where multiple UAVs provide  energy supply and communication services to IoT devices. They used a multi-UAV DQN based approach to improve throughput by jointly adjusting UAVs’ path design and time resource assignment. They follow an independent learner approach without cooperation between UAVs. In \cite{cruise}, the authors consider long-term, long-distance sensing tasks in a smart city scenario  where UAVs make decisions based on DQN for energy-efficient data collection. An energy-saving DRL-based UAV control strategy is developed in \cite{coverage} to enhance energy efficiency and communication coverage. They used deep deterministic policy gradient (DDPG) method and take into account communications coverage, fairness, energy consumption, and connectivity.
In \cite{multi-uav}, the dueling DQN is employed to adjust the UAV deployment in the multi-UAV wireless networks so that downlink capacity is to be enhanced while covering all ground terminals. They modeled the problem as a constrained MDP problem. 

The MARL framework is developed in \cite{resourceallocation} to investigate the dynamic resource allocation problem in UAV networks. A Q-learning based algorithm is developed to enhance the long-term rewards where each UAV runs Q-learning algorithm and automatically selects its communication mode, power levels and sub-channels in concurrent manner. \cite{spectrum} studies spectrum sharing among a network of UAVs. A relaying service is realized by team of UAVs to serve primary users on the ground aiming to gain spectrum access consequently. The gained spectrum  belongs to not only UAV relay, but also other UAVs that perform the sensing task. The problem is formulated as deterministic MMDP and distributed Q-learning is utilized to solve it. \cite{interference-aware} develops the DRL  algorithm  based  on  echo  state network  cells to find an interference-aware path and allocate resources to the UAVs. The developed scheme reduces wireless latency and improves energy efficiency. The work in \cite{trajectorydesign} adjusts trajectory and power control in multiple UAVs scenarios to enhance the users’ throughput and satisfying the users’ rate requirement. 
\section{DRL-aided flight resource allocation using mean field game} \label{sub:1}
In \cite{chen2020mean}, the authors explore energy-efficient control strategies for UAVs that provide fair communication coverage for ground users. The UAV control problem is modeled as an MFG and a mean-field TRPO algorithm is studied to design the UAVs' trajectories. In \cite{li2020downlink}, the authors apply the MFG theory to the downlink power control problem in ultra-dense UAV networks to improve the network's energy efficiency. Due to the complexity of the MFG, a DRL-MFG algorithm is developed to learn the optimal power control strategy. \cite{service} studies the task allocation in cooperative mobile edge computing and a mean field guided Q-function is formulated to reduce the network latency. MFG and DRL are integrated to guide the learning process of DRL according to the equilibrium of MFG. In \cite{sun2020joint}, the authors model the trajectory planning and power control for heterogeneous UAVs as an MFG, aiming to reduce energy consumption. A mean field Q-learning is studied to find the optimal solution. In \cite{uavpositionm}, the authors study UAV-assisted ultra-dense networks, where each UAV can adjust its location to reduce the AoI. They formulate the problem as an MFG and apply a DDPG-MFG algorithm to find the mean field equilibrium. In \cite{li2020downlink}, downlink power control for a large number of UAVs is suggested to enhance the energy efficiency  by learning the optimal power control policy. MFG is used to model the power control problem of the UAV network, where each UAV tries to enhance the energy efficiency by adjusting its transmit power. Then, due to the complexity of solving the formulated MFG, an effective DRL-MFG algorithm is suggested to learn the optimal power control strategy. 

Although, DRL-based solutions are mainly used, the following works adopt numerical solutions. In \cite{xue2018adaptive}, the focus is on adaptive coverage problem in emergency communication system, where multiple UAV act as aerial base stations to serve randomly distributed users. The problem is formulated using discrete MFG, each UAV aims to reduce its flight energy consumption and increase the number of users it can serve. Finally, optimal control and state of each UAV are computed. In \cite{xu2018discrete}, a discrete MFG is formulated to address joint adjustment of power and velocity for a large number of UAVs that act as aerial base stations. Decentralized  control laws are developed, and mean field equilibrium is analyzed. In \cite{velocitycontrol}, the authors present an energy-efficient velocity control algorithm for a large number of UAVs based on the MFG theory. The velocity control of the UAVs is modeled using a differential game in which energy and delay are balanced by using an original double mixed gradient method. 
\section{DRL-aided flight resource allocation for data freshness} \label{sub:2}
In \cite{9750860}, the authors consider ground sensors with limited energy and apply airborne base stations to collect sensory data. Each UAV's task is decomposed into energy transfer and fresh data collection. A centralized multi-agent DRL based on DDPG is developed to adjust the UAV trajectories in a continuous action space, to reduce the AoI of the ground sensors. In \cite{9909005}, the authors study UAV-assisted sensor networks where multiple UAVs cooperatively conduct the data collection to reduce the AoI. The trajectory planning is formulated as a decentralized partially observable markov decision process (Dec-POMDP). A multi-agent DRL is studied to find the optimal strategy. In \cite{hu2019distributed} and \cite{hu2020cooperative}, the authors develop the trajectory planning for multiple UAVs that perform cooperative sensing and transmission, aiming to reduce the AoI. In \cite{9285215}, ground sensors sample and upload data in a UAV-assisted IoT network. PPO is used to explore the optimal scheduling policy and altitude control for the UAV to reduce the AoI. In \cite{sun2021aoi}, a data collection scheme characterized by AoI and energy consumption in a UAV-assisted IoT network is investigated. The average AoI, and energy consumption of propulsion and communication are reduced by adjusting the UAV flight speed, hovering waypoints, and bandwidth allocation for data collection using a TD3-based approach.

 Although DDPG and PPO are used to adjust continuous and discrete actions to reduce AoI, the following works use DQN to adjust discrete actions. In \cite{9815722}, the authors investigate UAV-assisted IoT networks where multiple UAVs relay data between sensors and base station. A DQN-based trajectory planning algorithm is presented to reduce the AoI. In \cite{abd2019deep}, ground sensors with limited energy are used to observe various physical processes in the context of a UAV-assisted wireless network. The trajectory and scheduling policy are adjusted to reduce the weighted sum of AoI, and a DQN-based solution is applied to obtain the best strategy.  In \cite{zhou2019deep}, trajectory planning of the UAV is performed to reduce the AoI in a UAV-assisted IoT network. The problem is formulated as an MDP, and a DQN-based algorithm is studied to find the optimal trajectories of the UAV. In \cite{tong2020deep}, a UAV-assisted data collection for ground sensors is studied, where the UAV with limited energy is dispatched to collect sensory data. The UAV's trajectory is adjusted to reduce the average AoI and keep the packet loss rate low. The trajectory planning is formulated as an MDP while DQN is applied to design the UAV's trajectory. In \cite{liu2021average}, a UAV-assisted wireless network with an energy supply is used, where the UAV performs wireless energy transmission to ground sensors, and the sensors transmit data to the UAV using the harvested energy. A DQN-based trajectory planning algorithm is presented to reduce the average AoI by adjusting the trajectory, transmission schedule, and harvested energy. 
\section{Research Opportunity}
The works by \cite{obdrl} and \cite{kaicollection} address the optimization of velocity control and data collection schedules to minimize packet loss. However, these works are formulated for a single agent scenario. In contrast, our proposed approach, MADRL-SA, differs from the MARL framework introduced by Cui \cite{resourceallocation}. In MARL, UAVs operate based on an independent learner paradigm, whereas MADRL-SA promotes cooperation among UAVs to minimize packet loss. Additionally, MADRL-SA is specifically designed for practical scenarios and utilizes DQN, unlike MARL, which relies on Q-learning. The work by  \cite{dqn} adopts a single UAV approach, while MADRL-SA adopts a multi-UAV approach, offering advantages in terms of scalability and robustness. Our focus is on minimizing packet loss and providing velocity control, whereas the work by \cite{cruise}. (2017) prioritizes energy efficiency while neglecting velocity control. Furthermore, in the reviewed literature, the UAVs act independently without any explicit strategy for collaboration among them.

The existing literature in the fields of UASNets and DRL has yielded promising outcomes in tackling various challenges. Nevertheless, based on our current knowledge, no work has specifically focused on jointly optimizing cruise control and communication scheduling in the presence of multiple UAVs using DRL techniques. This presents an intriguing opportunity to explore innovative approaches to tackle this intricate problem.

Most of the works in Section  \ref{chap3:sec2} formulate MFGs to address energy efficiency in UASNets. For example, \cite{uavpositionm} propose an MFG formulation to minimize AoI and suggest the use of DDPG-MFG in a continuous action space to find the optimal solution. On the other hand, the works in Section \ref{sub:2} investigate resource allocation to reduce AoI, however, the actions are adjusted either in continuous or discrete action spaces. For instance, \cite{9285215} formulate a resource allocation problem to reduce AoI and employ PPO in a discrete action space to find the optimal solution.

Existing literature in the field of UASNets and AoI has shown promising results in addressing various challenges related to trajectory optimization and communication scheduling. However, most of the existing work focuses on single UAV scenarios, where actions are optimized in either continuous or discrete action spaces. On the other hand, research on multi-UAV systems using MFG formulations primarily targets energy efficiency. This presents an opportunity to explore novel approaches that utilize MFG formulations and optimize actions in mixed-action spaces to minimize AoI.

This thesis addresses the problem of joint velocity control and data collection scheduling in UASNets by formulating it as an MMDP to minimize overall packet loss caused by buffer overflow and channel fading. To handle the large state and action spaces, we propose MADRL-SA, which is based on DQN and enables the optimization of ground sensor selection, UAVs’ patrol velocity, and modulation scheme. 
Additionally, collaboration among UAVs is facilitated by allowing each ground sensor to maintain a history of UAV visits and share this information with other UAVs. Furthermore, this thesis formulates cruise control for multiple UAVs based on MFG to minimize the average AoI. We introduce MF-HPPO as a method to optimize the actions of UAVs in a mixed discrete and continuous action space. To capture temporal dependencies in the cruise control problem, we leverage an LSTM layer. By adopting these approaches, we aim to enhance the performance and efficiency of UASNets.

	   \clearemptydoublepage
           \chapter{Joint communication scheduling and velocity control in UAVs-assisted sensor networks: A deep reinforcement learning approach}
In this chapter, we address the joint optimization of communication scheduling and velocity control for multiple UAVs in UASNets. We formulate this problem as an MMDP aiming to minimize packet loss caused by buffer overflows and communication failures. The MMDP network state comprises battery levels, data queue lengths of ground sensors, channel conditions, visit times, and waypoints along the trajectories of the UAVs. UAVs take actions to schedule ground sensors for data transmissions, determine modulation schemes, and adjust patrol velocities. Ground sensors record and share visit times with UAVs as evidence of other UAVs' communication schedules. The rest of this chapter is organized as follows.  Section \ref{chap3:sec1} dedicates to the problem statement, where the system model is presented and the joint optimization of the velocity control and communication schedule is formulated. In Section \ref{chap3:sec2}, multi-UAV DQN is developed and a new MADRL-SA scheme is designed to optimize the decision process of the MMDP, thereby optimizing the patrol velocities as well as the transmission schedule of the ground sensors. Performance evaluation is presented in Section \ref{chap3:sec3}. This paper is concluded in Section \ref{chap3:sec4}.

\section{Problem Statement} \label{chap3:sec1}

\subsection{System Model}
\label{system model}
The network contains \emph{$J$} ground sensors and \emph{$I$} UAVs. Our study focuses on the joint velocity control and communication scheduling under preconfigured UAV trajectories. The UAVs fly along pre-determined trajectories which consist of a large number of waypoints to cover all the ground sensors in the field. The trajectories of the UAVs can be predesigned according to the required network capacity \cite{choi2014energy}, coverage \cite{dqn}, or the UAVs' propulsion energy consumption \cite{zeng2017energy}. The optimization of UAV trajectories has been widely studied in the literature \cite{9152044}, \cite{9154432}, \cite{9209079}. The proposed MADRL-SA is generic to any given trajectory.
\begin{table}[ht]
\centering
\caption{Notation and Definition}
\begin{tabular}{|c|c|} 
 \hline
 Notation & Definition \\  
 \hline
 \emph{$J$} & number of ground sensors  \\ 
 \hline
 \emph{$I$} & number of UAVs \\
 \hline
 $a_{u}^{t-1}$ & past actions of other UAVs on a ground sensor \\
 \hline
 $a_i$ & action of UAV i \\
 \hline
 $S_{\alpha,i}$ & state of UAV i \\
 \hline
 $S_{\beta,i}$ & next state of UAV i \\
 \hline
 $P_j^i(t)$& transmit power between device j and UAV i \\
 \hline
 $h_j^i(t)$& channel gain between device j and UAV i \\
 \hline
 $\zeta_i(t)$& location of the UAV on its trajectory\\
 \hline 
 $v(t)$& velocity of the UAV\\
 \hline
 $v_{max},v_{min} $& the maximum and minimum velocity of the UAV\\
 \hline
 $e_j(t)$& battery level of device j\\
 \hline
 $q_j(t)$& queue length of device j \\ 
 \hline
 $TVR_p$ &Time of each visiting record \\
 \hline
 \emph{$D$} & maximum queue length of ground sensor\\
 \hline
 $\phi_j (t)$ & modulation scheme of device j\\
 \hline
 $\gamma$ & discount factor for future states\\
 \hline
 $\theta$&learning weight in deep Q-network\\
 \hline
\end{tabular}
\label{table:1}
\end{table}   
The channel coefficient between the UAV $\emph{$i$}$ $(\in [1,\emph{$I$}])$ and device $\emph{$j$}$ $(\in [1, \emph{$J$}])$ at \emph{$t$} is $h_j^i(t)$, which can be known by channel reciprocity. The modulation
scheme of device \emph{$j$} at  \emph{$t$} is denoted by $\phi_j (t)$. In particular, $\phi_j (t)$= 1, 2, and 3 indicates binary phase-shift keying (BPSK), quadrature-phase shift keying (QPSK), and 8 phase-shift keying (8PSK), respectively, and $\phi_j (t)$ $\geq$ 4 provides $2^{\phi_j (t)}$ quadrature amplitude modulation (QAM).

Let $h_j^i(t)$ denote channel gain between ground sensor \emph{$j$} and UAV \emph{$i$}. The transmit power of the ground sensor, denoted by $P_j^i(t)$, is\cite{ddpg}

\begin{equation}
 P_j^i(t)=\frac{\emph{$ln$}\frac{k_1}{\epsilon}}{k_2 {h_j^i(t)}^2}(2^{\phi_j (t)}-1)  
\end{equation}
where \emph{$k_1$} and \emph{$k_2$} are channel constants, and $\epsilon$ denotes the required bit error rate (BER) of the channel.
We consider that UAV \emph{$i$} moves in low attitude for data collection, where the probability of LoS communication between UAV \emph{$i$} and ground sensor \emph{$j$} can be 
\begin{equation}
    Pr_{LoS}(\varphi_j^i)=\frac{1}{1+a exp(-b[\varphi_j^i-a])}
\end{equation}
where \emph{$a$} and \emph{$b$} are constants, and  $\varphi_j^i$ denotes the elevation angle between  UAV \emph{$i$} and ground sensor
\emph{$j$}. Furthermore, path loss of the channel between  UAV \emph{$i$} and device \emph{$j$} can be obtained by 
\begin{equation}
\gamma_j^i=Pr_{LoS}(\varphi_j^i)(\eta_{LoS}-\eta_{NLoS})+\\20 log(r \sec(\varphi_j^i))+20 log(\lambda)+ 20 log(\frac{4\pi}{v_c})+\eta_{NLoS}  
\end{equation}
where \emph{$r$} denotes the radius of the radio coverage of UAV \emph{$i$}, \emph{$\lambda$} is the carrier frequency, and \emph{$v_c$} is the speed of light. $\eta_{LoS}$ and $\eta_{NLoS}$ represent the excessive path losses of LoS or non-LoS, respectively\cite{lap}. Please See Appendix A.
\subsubsection{Communication Protocol}
Fig. \ref{fig:communication} shows the data collection protocol for the UASNets. Specifically, the proposed MADRL-SA operates onboard at the UAVs to determine their velocities and sensor selection and allocate the modulation scheme for the selected sensors. The details of MADRL-SA will be provided in the next section. Next, the UAV  broadcasts a short beacon message which contains the ID of the selected sensor. Upon the receipt of the beacon message, the selected sensor transmits its data packets to the UAV, along with the state information of $e_j(t)$, $q_j(t)$, and $TVR_p$  in the control segment of the data packet. Once the data is correctly received, the UAV sends an acknowledgment to the ground sensor.  
\begin{figure*}[ht]
        \centering
    	\includegraphics[ width=5.5in, height =3in]{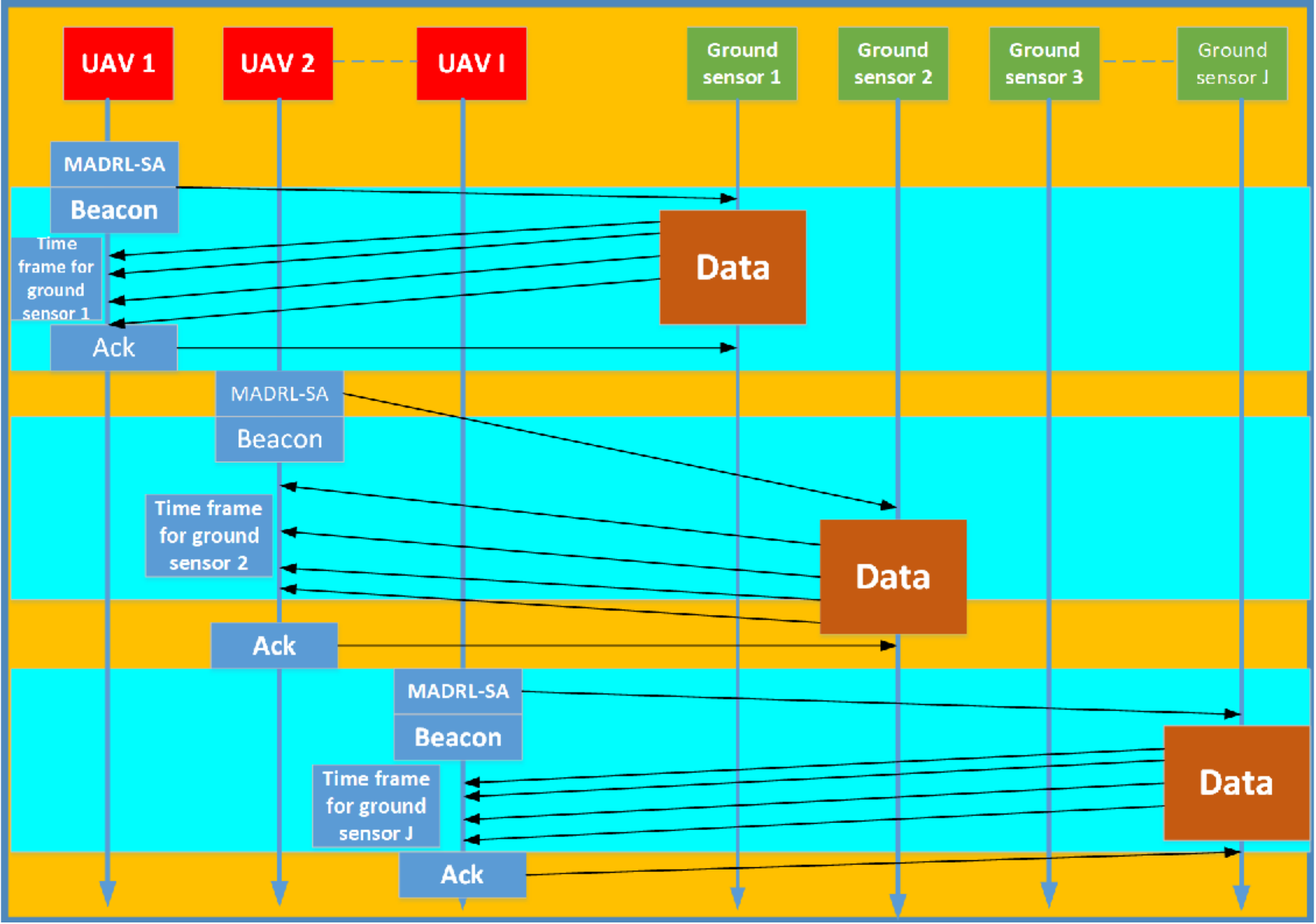}
    	\caption{Data communication protocol for UASNets. MADRL-SA conducts velocity determination, sensor selection, and modulation scheme allocation in each communication frame}
    	\label{fig:communication}
\end{figure*}
\subsection{Problem Formulation}
\label{mmdp}
In this section, we present the problem formulation.
\\

\subsubsection{Optimization Formulation}
Let $\kappa _{j}^{i}(t)$ be the binary indicator of ground sensor $j$ being selected by UAV $i$ for data transmission at time $t$. If ground sensor $j$ is scheduled by  UAV $i$ at time $t$, $\kappa_{j}^{i}(t) = 1$; otherwise, $\kappa_{j}^{i}(t)=0$. The joint optimization of UAV velocity and communication schedule aims to minimize the packet loss of all the ground sensors, as given by 
\\  Optimization problem:
    \begin{align}
    \label{eq:objective}
    &\min_{\kappa_j^i(t),v_i(t),P_{j}^{i}(t)} \sum_{i=1}^{I} \sum_{j=1}^{J} f_{ij}(\kappa_j^i(t),v_i(t),P_{j}^{i}(t)) + \sum_{j=1}^J g_j(\kappa_j^i(t)) \nonumber \\
    &\text{subject to:} \nonumber
\end{align}
   \begin{equation}
    \label{eq:power}
    0 \leq P_{j}^{i}(t)\kappa_j^i(t) \leq P_{max},
\end{equation}
   	
where 
\\
\\
\begin{equation}
f_{ij}(\kappa_j^i(t),v_i(t),P_{j}^{i}(t)) = \\
\begin{cases}  
1, & \text{if } (\kappa_j^i(t)=1) \text{ \& } (h_{j}^{i}(t) \le h_{th}) \text{ \& } (v_i(t) \le v_{max}) ; \\
0, & \text{otherwise,}
\end{cases}
\end{equation}
\\
and
\begin{equation}
g_j(\kappa_j^i(t)) =
\begin{cases}  
1, & \text{if } (q_j(t)>D) \text{ \& } (\kappa_j^i(t)=0); \\
0, & \text{otherwise,}
\end{cases}
\end{equation}

Constraint (\ref{eq:power}) ensures that the transmit power of the scheduled ground sensor does not exceed the maximum transmit power $P_{max}$.
\subsubsection{MMDP Formulation}
MMDP can be  defined by the tuple
$\{I, \{S_{\alpha,i}\}, \{a_i\}, C\{S_{\beta} \mid S_{\alpha}, a\}, \Pr\{S_{\beta} \mid S_{\alpha}, a\}\}$

       \begin{enumerate}
	    \item \emph{I} is the number of agents, i.e., UAVs.
	    \item $S_{\alpha,i}$ is the network state observed by agent $i$ ($i\in I$). $S_{\alpha,i}$ comprises: channel quality $h_j^i(t)$, battery level $e_j(t)$, queue length $q_j(t)$, visit time $TVR_p$, and the location of UAV $\zeta_i(t)$, i.e., $S_{\alpha,i}=\{(h_j^i(t), e_j(t), q_j(t), TVR_p, \zeta_i(t)), i=1,2,\dots, I\}$.

	    In particular, each ground sensor maintains a list of visiting time of the agents. Joint state of all the agents is denoted  $S_{\alpha}$, where $S_{\alpha}$=$S_{\alpha,1}$$\times......\times$$S_{\alpha,I}$. 
	    \item $a_i$ represents the action of agent $i$. $a_i$ is to schedule one sensor to transmit data to the UAV, determine the modulation and the instantaneous patrol velocity of the UAV, i.e., $a_i$=$\{(j, \phi_j(t), v(t)), i=1,2,....I\}$. Joint action $a$ which consists of the actions of all the agents is   $a$=$a_1$$\times......\times$$a_I$. The size of action space is $J\Phi\mid v(t)\mid$, where $\Phi$ is the highest modulation order and  $\mid v(t)\mid$ stands for the cardinality of the set $[v_{min},v_{max}]$.
	    \item C\{$S_{\beta}$$|$$S_{\alpha}$, $a$\} is the network cost yielded when joint action $a$ is taken at joint state $S_{\alpha}$ and the following joint state changes to $S_{\beta}$. The network cost is the packet loss of the	ground sensors.
	   \item $\Pr\{S_{\beta} \mid S_{\alpha}, a\}$ denotes the transition probability from joint state $S_{\alpha}$ to joint state $S_{\beta}$ when joint action $a$ is taken.

	\end{enumerate}

 \subsubsection{Transition Probability} \label{sub:100}
The transition probability of the MMDP, from $S_{\alpha}$ to $S_{\beta}$ can be given by 
       \begin{multline} 
	\label{eq:444}
	    \Pr\{S_{\beta} \mid S_{\alpha}\} = \prod_{i=1}^I \left(\Pr\{(e_{\beta,j}, q_{\beta,j}, h_{\beta,j}, \zeta_{\beta,j}) \mid (e_{\alpha,j}, q_{\alpha,j}, h_{\alpha,j}, \zeta_{\alpha,j}), j \in a_i\}_i \right) \\ \times \prod_{k=1}^K \left(\Pr\{(e_{\beta,k}, q_{\beta,k}, h_{\beta,k}, \zeta_{\beta,k}) \mid (e_{\alpha,k}, q_{\alpha,k}, h_{\alpha,k}, \zeta_{\alpha,k}), k \ne a_i; i \in [1,I]\}_i\right)
\end{multline}

Specifically, the state transition probability presented in  (\ref{eq:444}) consists of two parts. The first part, i.e.,
$\Pr\{(e_{\beta,j}, q_{\beta,j},h_{\beta,j},\zeta_{\beta,j})|(e_{\alpha,j},q_{\alpha,j},h_{\alpha,j},\zeta_{\alpha,j}),j \in a_i\}$is the state transition probability from $S_\alpha$ to $S_\beta$ in terms of the selected ground sensor ($j \in a_i $). Let K denote the total number of unselected ground sensors. The second part, i.e., 

$\Pi_{k=1}^K \Pr\{(e_{\beta,k},q_{\beta,k},h_{\beta,k},\zeta_{\beta,k}) |(e_{\alpha,k},q_{\alpha,k},h_{\alpha,k},\zeta_{\alpha,k},k \ne a_i; i \in [1,I]\}$

is the probability from $S_\alpha$ to $S_\beta$ in terms of the unselected ground sensors, where $k \ne a_i;i\in[1,I]$ indicates the sensors that are not selected by any of the $I$ agents.
\par
Let $d_{i,j}$ denote the distance between ground sensor \emph{j} and UAV \emph{i}, $v(t)$ is velocity of the UAV , $R(t)$ is the data rate of the ground sensor and $\lambda$ is the packet arrival probability. The state transition probability of the selected sensor $j$, which is specified in (\ref{eq:5}), depends on the following possible transitions.

\begin{enumerate}
    \item Packet transmission is successful due to the good channel quality, i.e., $h_{\beta,j}>h_{\alpha,j}$ and low velocity. There is no packet arrival, the data queue of the selected node decreases, i.e.,  $q_{\beta,j}=q_{\alpha,j}-1$. The state transition probability is $(1-\epsilon)^{\frac{2d_{i,j}R(t)}{v(t)}}(1-\lambda)$.\\
    \item   Packet transmission is failed due to the poor channel quality, i.e., $h_{\beta,j}<h_{\alpha,j}$ and high velocity. A new data packet is generated and buffered, the data queue of the selected node increases, i.e., $q_{\beta,j}=q_{\alpha,j}+1$. The state transition probability is $(1-(1-\epsilon)^{\frac{2d_{i,j}R(t)}{v(t)}})\lambda$. \\
    \item Packet transmission is successful due to the good channel quality, i.e., $h_{\beta,j}>h_{\alpha,j}$ and low velocity. A new data packet is generated and buffered, the data queue of the selected node remains unchanged, i.e., $q_{\beta,j}=q_{\alpha,j}$. The state transition probability is $(1-\epsilon)^{\frac{2d_{i,j}R(t)}{v(t)}}\lambda$. \\
    \item Packet transmission is failed due to the poor channel quality, i.e., $h_{\beta,j}<h_{\alpha,j}$ and high velocity. There is no packet arrival, the data queue of the selected node remains unchanged, i.e.,  $q_{\beta,j}=q_{\alpha,j}$. The state transition probability is $(1-(1-\epsilon)^{\frac{2d_{i,j}R(t)}{v(t)}})(1-\lambda)$. \\
    
\end{enumerate}

\begin{figure*}

\begin{multline}
\label{eq:5}
\Pr\{(e_{\beta,j}, q_{\beta,j}, h_{\beta,j}, \zeta_{\beta,j}) \mid (e_{\alpha,j}, q_{\alpha,j}, h_{\alpha,j}, \zeta_{\alpha,j}), j \in a_i\} = \\
\begin{cases}
(1-\epsilon)^{\frac{2d_{i,j}R(t)}{v(t)}}(1-\lambda) & \text{if } e_{\beta,j} = e_{\alpha,j} - \Delta e \text{ and } q_{\beta,j} = q_{\alpha,j} - 1 \\ & \text{and } h_{\beta,j} > h_{\alpha,j} \\
(1-(1-\epsilon)^{\frac{2d_{i,j}R(t)}{v(t)}})\lambda & \text{if } e_{\beta,j} = e_{\alpha,j} - \Delta e \text{ and } q_{\beta,j} = q_{\alpha,j} + 1 \\ & \text{and } h_{\beta,j} < h_{\alpha,j} \\
(1-\epsilon)^{\frac{2d_{i,j}R(t)}{v(t)}}\lambda & \text{if } e_{\beta,j} = e_{\alpha,j} - \Delta e \text{ and } q_{\beta,j} = q_{\alpha,j} \\ & \text{and } h_{\beta,j} > h_{\alpha,j} \\
(1-(1-\epsilon)^{\frac{2d_{i,j}R(t)}{v(t)}})(1-\lambda) & \text{if } e_{\beta,j} = e_{\alpha,j} - \Delta e \text{ and } q_{\beta,j} = q_{\alpha,j} \\ & \text{and } h_{\beta,j} < h_{\alpha,j}
\end{cases}
\end{multline}
\begin{equation}
\label{eq:6}
\Pr\{(e_{\beta,k},q_{\beta,k},h_{\beta,k},\zeta_{\beta,k})|(e_{\alpha,k},q_{\alpha,k},h_{\alpha,k},\zeta_{\alpha,k},k \ne a_i; i \in [1,I]\}= \begin{cases}  
\lambda    &    \text{if  $e_{\beta,k}=e_{\alpha,k}$ and $q_{\beta,k}=q_{\alpha,k}+1$}\\
1-\lambda    & \text{if $e_{\beta,k}=e_{\alpha,k}$ and $q_{\beta,k}=q_{\alpha,k}$}\\
0          & \text{otherwise}

\end{cases} 
\end{equation}

\end{figure*}
Due to the packet transmission, the battery level of the selected sensor decreases by $\Delta e$.\\
(\ref{eq:6}) corresponds to the unselected sensors with two different cases.
The first case corresponds to the case when queue of the ground sensor increases, i.e., $q_{\beta,k}=q_{\alpha,k}+1$  due to a new packet arrival, i.e., $\lambda$. The second case gives that the data queue remains unchanged, i.e., $q_{\beta,k}=q_{\alpha,k}$ since there is no packet arrival, i.e., $(1-\lambda)$.
\par
By solving the formulated MDP, e.g., by using dynamic programming techniques, the optimal solution with complete states could be achieved, which could be used for performance benchmarking in multi-UAV-assisted wireless sensor networks. Unfortunately, dynamic programming (and the MDP formulation) suffers from the well-known curse-of-dimensionality, and incurs a prohibitive complexity and intractability, which can be noted in Appendix B. Please See Appendix B.
\section{Proposal}  \label{chap3:sec2}
\begin{algorithm} [ht]
\caption{MADRL-SA}
\SetAlgoLined
 \textbf{1.Initialize:}
 \newline
  Randomly initialize the networks \\
 $Q_i\{S_{\beta,i}\mid S_{\alpha,i},a_i,a_{u}^{t-1};\theta^{Q_i}\}$ with $\theta^{Q_i}$
 \newline
 Initialize target networks $Q_i^{\prime}$ with weights $\theta^{Q_i^\prime}=\theta^{Q_i}$
  $\forall i\in(1,I)$
 
      \textbf{2.Learning:}
      
        \For {$episode=1$ to M }{
    
           Obtain state $S_{\alpha,i}$
                    
            \For {$t=1$ to T }{

                     \textbf{if}( Probability $\varepsilon$){
                     
                      Select a random action $a_i$
                                                          }
                  \newline                                        
                  \textbf{else}
                  
                  { 
                     $a_i=argmin_{a_i} Q_i\{S_{\beta,i}\mid S_{\alpha,i},a_i,a_{u}^{t-1};\theta^{Q_i}\}$
                   }
                   
                   \textbf{end} 
                   Execute action $a_i$ in the environment
                  \newline
                   Receive the visiting record
                    
                    \For {$p=1$ to I }{
                     
                     \textbf{if}(i==p)
                     \newline
                     $\delta[p]$=t
                     \newline
                     \textbf{else}
                     \newline
                        $\delta[p]$=t - $TVR_p$
                     
                     \textbf{end}}
                  
                   Obtain the cost function
                   $C_{t,i}=\{S_{\beta,i}\mid S_{\alpha,i},a_i,a_{u}^{t-1}\}$ and the next state $S_{\beta,i}$ at $t+1$
                   \newline 
                    Store Transition $(S_{\alpha,i},S_{\beta,i},a_i,C_{t,i})$
                   
                    Sample random minibatch $(S_{\alpha,b},S_{\beta,b},a_b,C_{t,b})$
                 
                $y_i=C\{S_{\beta,b} \mid S_{\alpha,b},a_b,a_{ub}\}+ \gamma min_{a_b^\prime} Q_i^{\prime}\{S_{\beta,b^\prime} \mid S_{\beta,b},a_b^\prime, a_{ub}^\prime;\theta^{Q_i^\prime}\}$
                \newline
                 Derive the loss function
              \newline
              $\Gamma(\theta_i^{Q})=y_i-Q_i\{S_{\beta,b}\mid S_{\alpha,b},a_b, a_{ub};\theta^{Q_i}\}$   
              \newline
               Update the target networks.
                \newline
              $\theta^{Q_i^\prime}=\theta^{Q_i}$
              $S_{\alpha}=S_{\beta}$
            }
          }
\end{algorithm}
  
\subsection{Proposed MADRL-SA} 
We present a multi-UAV version of DQN called MADRL-SA, MADRL-SA realizes cooperation between UAVs, by enabling them to learn the scheduling decisions of each other.

\begin{figure*}[ht] 
        \centering
        \captionsetup{justification=centering}
    	\includegraphics[ width=5in]{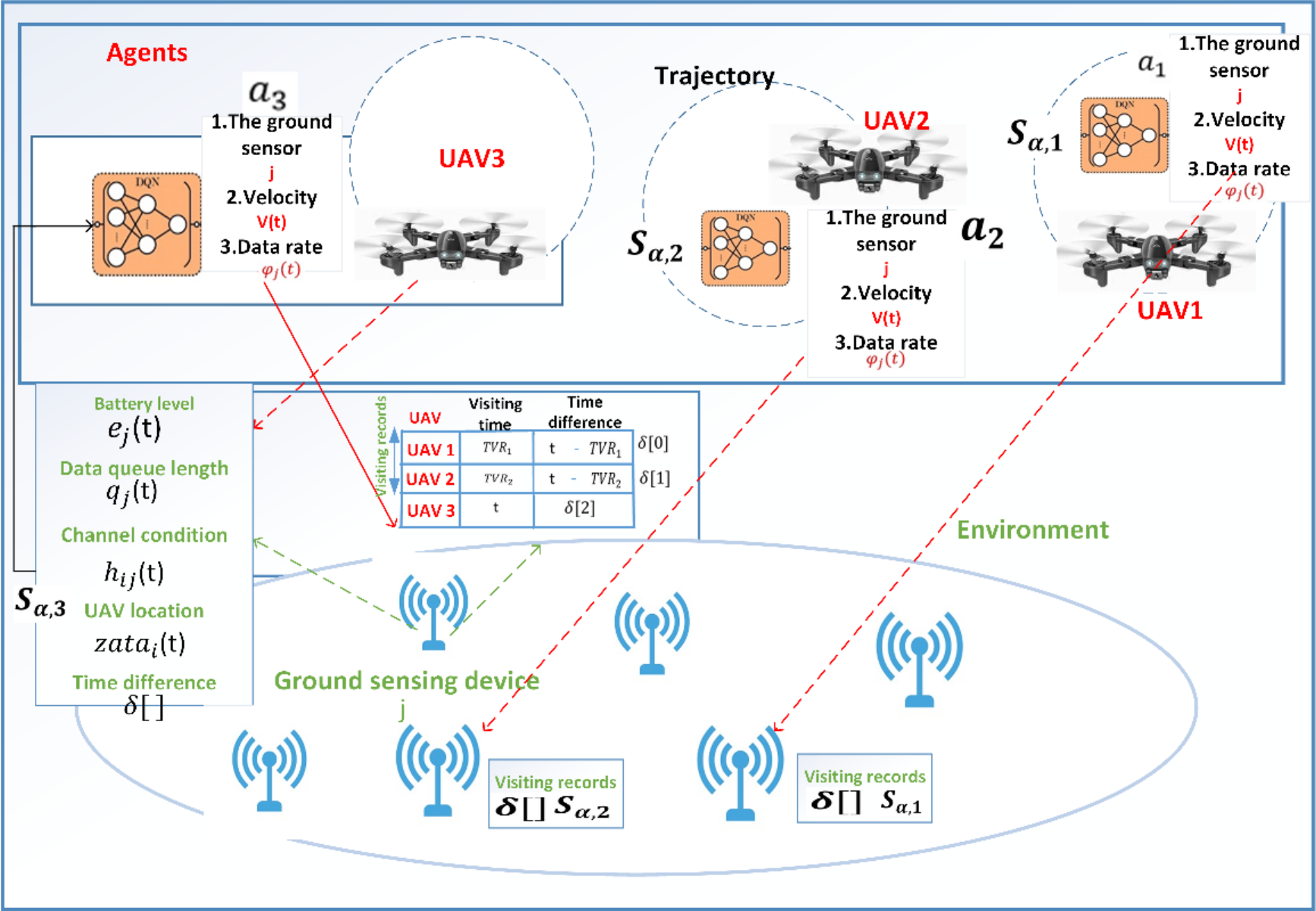}
    	\caption{Overview of MADRL-SA: UAVs observe the current environment state, follow their policy, and take actions}
    	\label{fig:UASNets}
\end{figure*}

According to Fig. \ref{fig:UASNets}, MADRL-SA  has three UAVs, and each UAV is equipped with a classical DQN algorithm and learns through interaction by environment.  As can be seen in Fig.\ref{fig:UASNets}, UAV 3 performs its action and schedules a ground sensor, then receives its visiting record and consequently calculates the time differences $\delta[]$ between its visiting time(t) and $TVR_p$. $\delta[]$ is augmented to state and utilized in the learning process. Therefore, each UAV learns to coordinate its action. The UAVs that visited the same ground sensor would learn to improve their scheduling process based on computed timing information. For example, if the computed time differences are large the UAV is encouraged to schedule the ground sensor for the next time. Overall, our goal is to allow different UAVs schedule different ground sensors (other ground sensors may have buffer overflow probability) and if a ground sensor recently visited by an UAV no other UAV visits that ground sensor. The proposed scheme is described in Algorithm 1, which optimizes the actions based on the multi-UAV DQN to solve the online resource allocation problem.

Overall, two separate Q-networks are maintained with each UAV, Q-network: $Q_i\{S_{\beta,i}\mid S_{\alpha,i},a_i, a_{u}^{t-1};\theta^{Q_i}$\} and target network: $Q_i^{\prime}\{S_{\beta,i^\prime} \mid S_{\beta,i},a_i^\prime, a_{u}^\prime;\theta^{Q_i^\prime}\}$, with weights $\theta^{Q_i}$ and $\theta^{Q_i^\prime}$ respectively. At first step, Q-network and  associated target of each UAV are initialized and then learning is ignited. Each UAV samples its state and computes its local state $ S_{\alpha,i}$ including $\delta[]$. Each UAV receives the local state $ S_{\alpha,i}$  and selects a random action with probability $\varepsilon$ or exploits its knowledge and produce its action. Each UAV executes the selected action and computes the vector of $\delta$ using \emph{t} and $TVR_p$;
then corresponding cost and next state including $\delta[]$ are sampled. Then the associated  transition $(S_{\alpha,i},S_{\beta,i},a_i,C) $ is stored. 
$\theta^{Q_i}$ is learned by sampling batches of  transitions from the replay memory and minimizing the squared temporal difference error:
\begin{equation}
   \Gamma(\theta^{Q_i})=y_i-Q_i\{S_{\beta,b}\mid S_{\alpha,b},a_b, a_{ub};\theta^{Q_i}\}   
\end{equation}
where
\begin{multline}
y_i=C\{S_{\beta,b} \mid S_{\alpha,b},a_b,a_{ub}\}+ \gamma min_{a_b^\prime} Q_i^{\prime}\{S_{\beta,b^\prime} \mid S_{\beta,b},a_b^\prime, a_{ub}^\prime;\theta^{Q_i^\prime}\}
\end{multline}
finally for each agent the parameters of a Q-network $\theta^{Q_i}$ 
 copied into those of target network $\theta^{Q_i^\prime}$ after a constant number of iterations. The proposed MADRL-SA can be readily repurposed to support different objective functions. For example, it can be potentially repurposed to maximize the energy efficiency, which is the ratio of network throughput to the energy consumption.
 
\subsection{Energy and Feasibility}
UAVs are becoming increasingly less restrictive in terms of energy due to new advancements of battery and energy harvesting technologies. For example, Atlantik Solar has developed an autonomous, solar-powered drone (UAV) that can fly up to 10 days continuously. A ground sensor can be equipped with solar panels, wind power
generators or other energy harvesting mechanisms to harvest renewable energy from ambient resources and recharge its battery.

The UAVs select the optimal sensors to transmit data and allocate their modulation schemes, by learning the states of the ground sensors. The selected sensor uses the allocated modulation to transmit data to the UAV, while updating the visiting time of the UAV. In particular, the historical record of the visiting time typically has a small size. Consider 100 UAVs, the size of the historical record at the sensor is just seven bits. The time for updating the record is negligible. Also, the sensors only need to synchronize with the UAVs the recent historical record of visits. The overhead is small. Therefore, the proposed deep reinforcement learning based data collection requires a small amount of computation at the sensors, which is feasible and practical in real-world UASNets
\subsection{Complexity of MADRL-SA}
The time complexity for training each network $Q_i$ that has $Z$ layers with $z_i$ neurons per layer is given by, 
\begin{equation}
    \mathcal{O}(MT \times (\Sigma_{i=1}^{Z-1}{z_i z_{i+1}}))
\end{equation}
where M is the number of episodes and T is the number of iterations. Therefore, the time complexity of MADRL-SA with I networks of $Q_i$
is given by 
\begin{equation}
    \mathcal{O}(I \times MT \times (\Sigma_{i=1}^{Z-1}{z_i z_{i+1}}))
\end{equation}
The case of an  equal number of neurons in each layer, the time complexity can be written as
\section{Evaluation} \label{chap3:sec3}
\subsection{Implementation of MADRL-SA}
J number of ground sensors are randomly deployed, where J increases from 20 to 120. Each ground sensor  has the maximum discretized battery capacity 50 Joules, the highest modulation = 5, and the maximum transmit power 100 milliwatts. For calculating $P_j^i(t)$ of the ground sensor, the two channel constants, $k_1$ and $k_2$ are set to 0.2 and 3, respectively. The required BER is 0.05, and the carrier frequency is 2000 MHz. $\varepsilon$ is set to $0.05$. However, the value of $\varepsilon$ can be configured based on the traffic type and quality-of-service (QoS) requirement of the user’s data, as well as the transmission capability of the UAV. Other simulation parameters are listed in Table \ref{table:222}. Moreover, the region of interest is set to be a square area with a size of 1000 x 1000 meters, where the ground sensors are distributed in the targeted region. MADRL-SA is implemented in Python 3.5 using Pytorch  (the Python deep learning library). A Lenovo Workstation running 64-bit Ubuntu 16.04 LTS, with Intel Core i5-7200U CPU @ 2.50GHz × 4 and 8 G memory is used for the PyTorch setup. DRL trains MADRL-SA for 1000 episodes. The discount factor and learning rate are set to 0.99 and 0.001, respectively. We use  2-layer fully connected neural network for each agent, which includes 400 and 300 neurons in the first and second layers, respectively. We utilize the rectified linear unit (ReLU) function for the activation function. The experience replay memory with the size of $10^6$ is created for each agent to store the learning outcomes in the format of a quadruplet \textless state, action, cost, next state\textgreater. The memory is updated by calling the function replay buffer$_i$.add((state, action, cost, next state)), and retrieves the experiences by using replay buffer$_i$.sample(batch size).
\begin{table}[ht]
\centering
\caption{PyTorch Configuration}
\begin{tabular}{|c|c|} 
 \hline
 Parameters & Values \\  
 \hline
 Number of ground sensors & 20-120  \\ 
 \hline
  Queue length & 40 \\
 \hline
 Energy levels & 50 \\
 \hline
 Discount factor & 0.99 \\
 \hline
 Learning rate & 0.001 \\
 \hline
 Replay memory size & $10^6$ \\
 \hline
 Batch size & 100 \\
 \hline
 Number of episodes & 1000 \\
 \hline
\end{tabular}
\label{table:222}
\end{table}

\subsection{Baseline Description}
For performance evaluation, the proposed MADRL-SA is compared with Random scheduling policy (RSA), Channel scheduling policy (CHSA) and DRL-SA \cite{dqn} algorithms. 
\begin{itemize}
    \item RSA randomly determines the velocities of the UAVs at each waypoint, and one of the ground sensors within the communication range of the UAV is randomly selected to transmit data. The velocity control and sensor selection are independent of the batteries, data queue lengths of the ground sensors, channel variation, and UAVs' positions.
    \item CHSA allows the UAVs to move with the minimum velocity and schedule the ground sensors based on their channel quality. Each UAV sends beacons along the trajectory. Based on the sensors' replies to the beacons, the UAV measures the channel gains. The ground sensor with the highest channel gain is selected to transmit.
    \item DRL-SA enables a single-agent DQN, where each UAV leverages DQN to learn the optimal velocity control and sensor selection strategy based on the data queue length, energy level, channel variation and UAV’s positions. The selection of the ground sensor, modulation scheme, and velocity of the UAV is jointly optimized (independently of the rest of the UAVs). 
\end{itemize}
\subsection{Performance Analysis of MADRL-SA}
\begin{figure} 
        \includegraphics[ width=3.8in, height =2in]{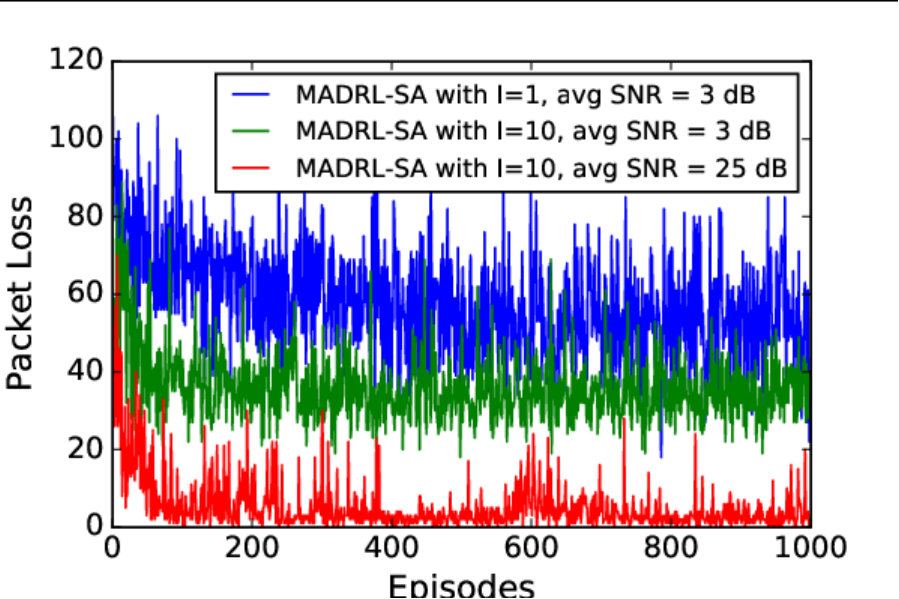}
            \Centering
    	\caption{Network cost at each episode of MADRL-SA with I = 10 and DRL-SA.
.}
    	\label{fig:agent}
\end{figure}

\begin{figure}
    	\includegraphics[ width=3.8in, height =2in]{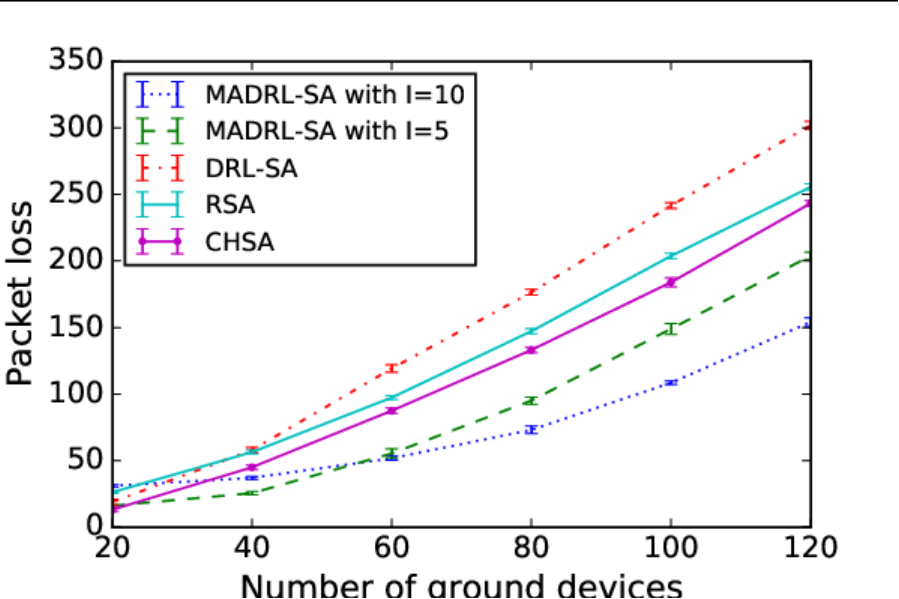}
            \Centering
    	\caption{Comparison of packet loss between MADRL-SA and the baselines in terms of ground sensors. }
    	\label{fig:network}
\end{figure}

\begin{figure}[ht] 
        \includegraphics[ width=3.6in]{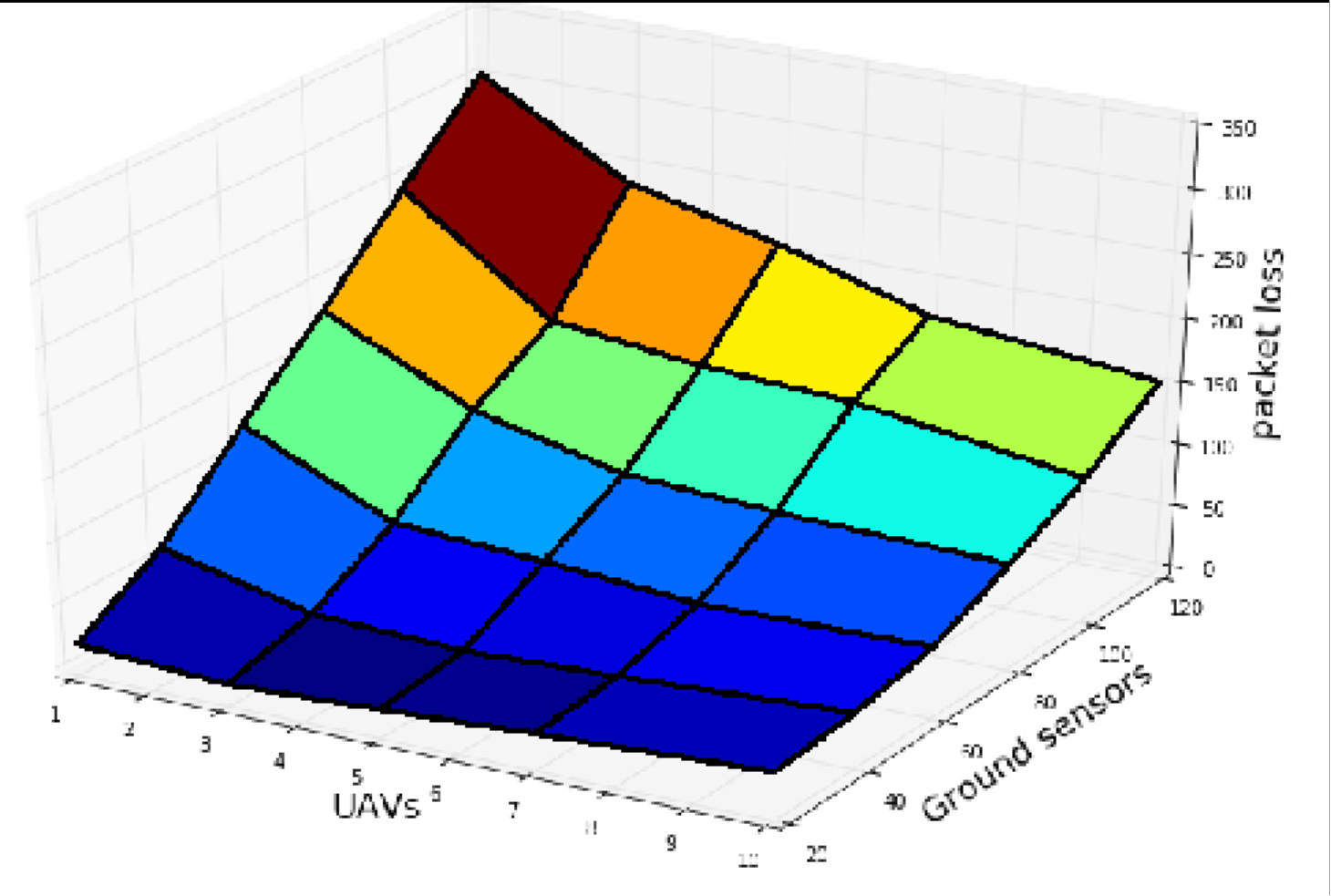}
            \Centering
    	\caption{Trade-off between the number of UAVs and ground sensors.}
    	\label{fig:packet}
\end{figure}
\begin{figure}[ht] 
        \includegraphics[ width=3.6in]{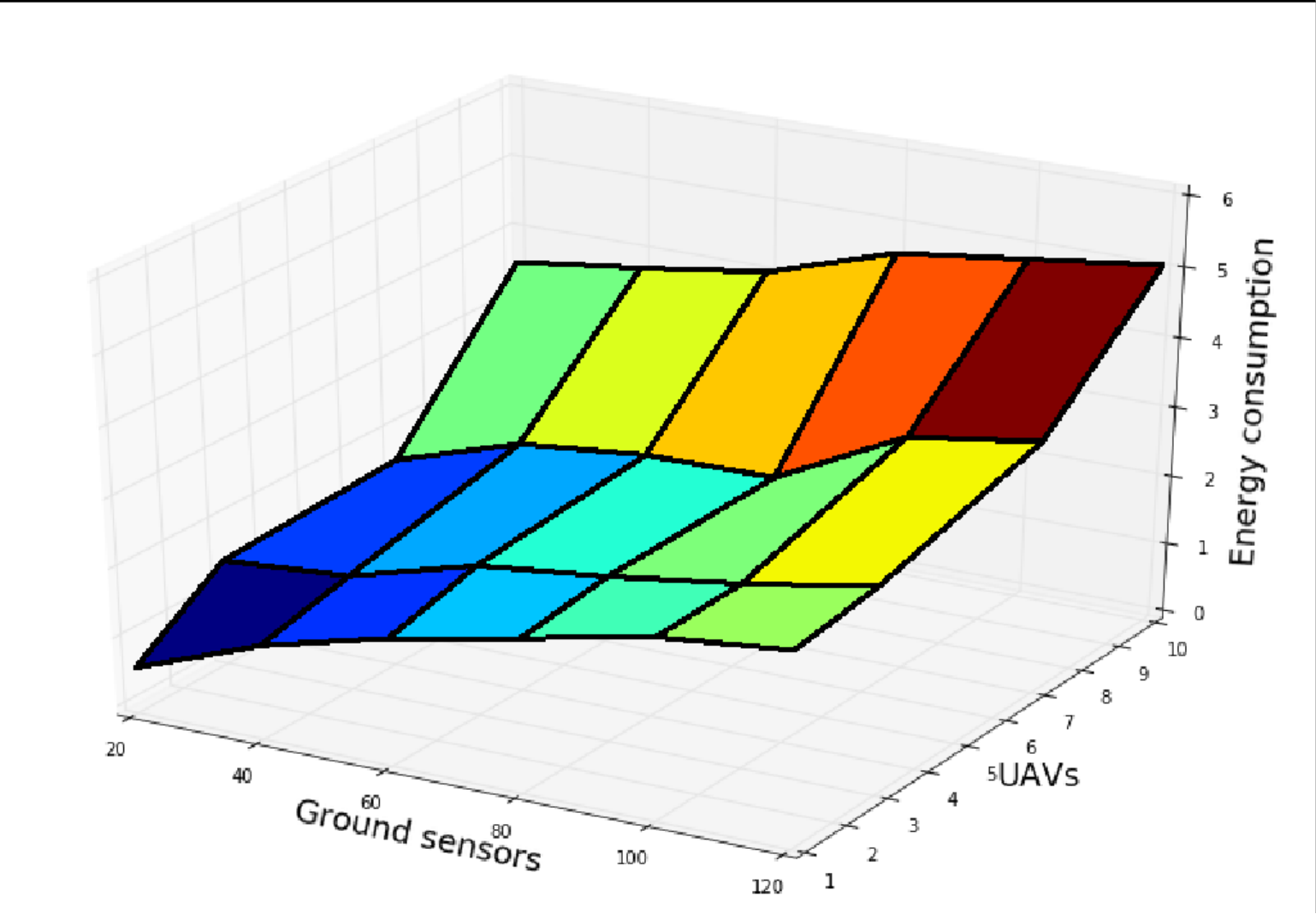}
            \Centering
    	\caption{Energy consumption of ground sensors.}
    	\label{fig:energy}
\end{figure}

\begin{figure*}[ht]
    \begin{subfigure}[b]{0.45\textwidth}
        \includegraphics[ width=3in, height =1.60in]{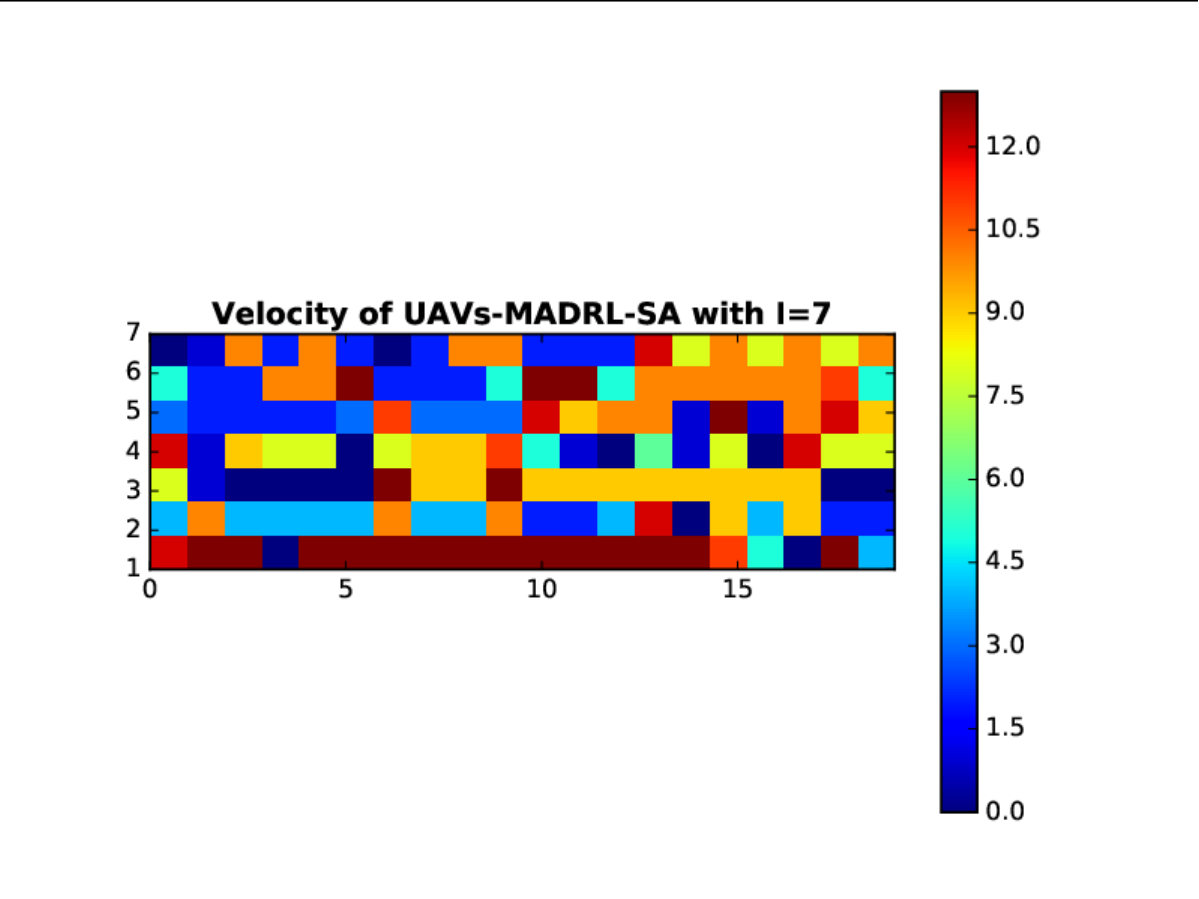}
        \caption{}
        \label{fig:velocity20}
    \end{subfigure}
    \hfill
    \begin{subfigure}[b]{0.45\textwidth}
        \includegraphics[ width=3in, height =1.60in]{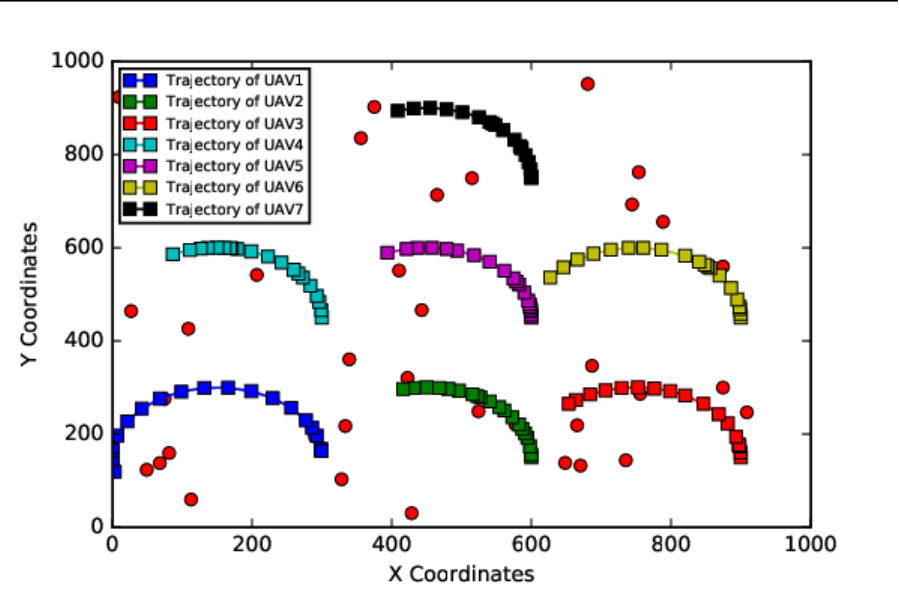}
        \caption{}
        \label{fig:trajectory20}
    \end{subfigure}

    \begin{subfigure}[b]{0.45\textwidth}
        \includegraphics[ width=3in, height =1.60in]{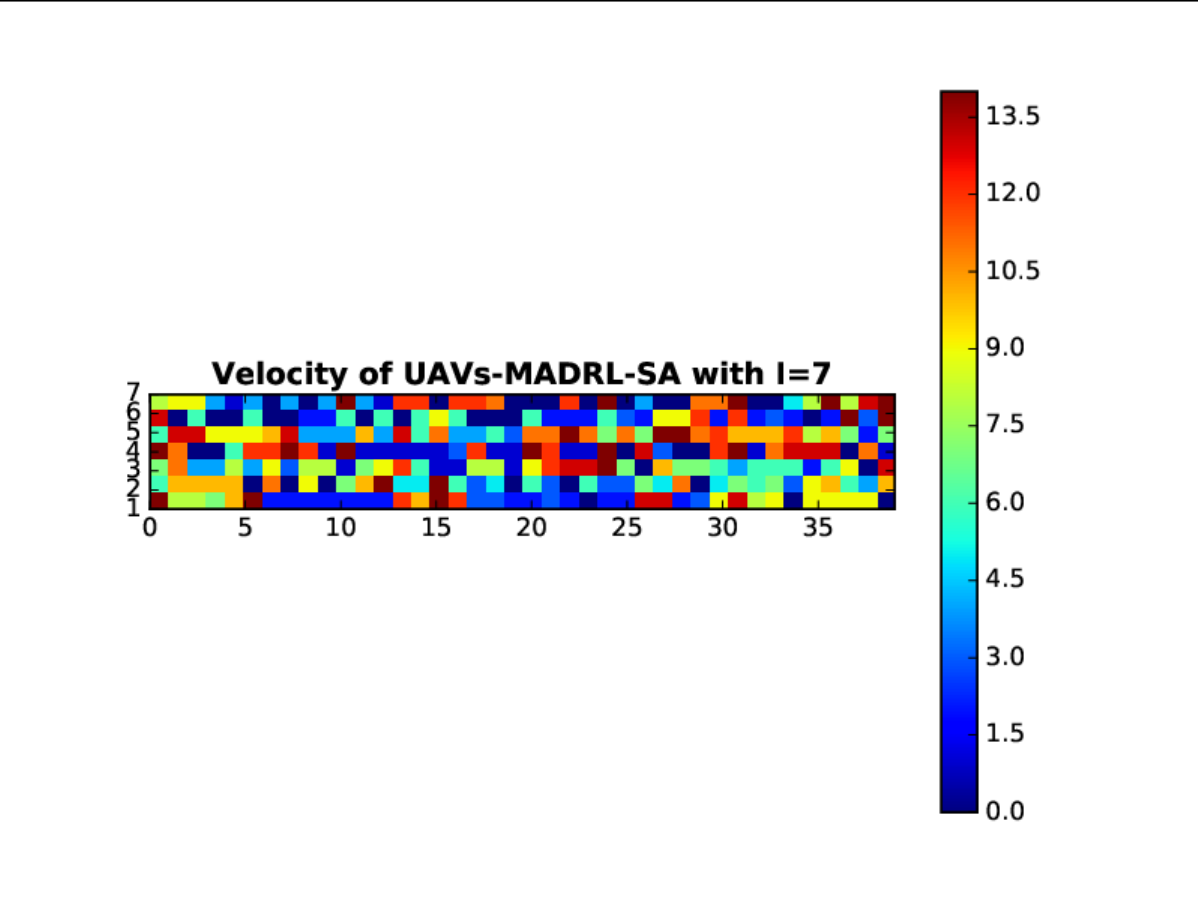}
        \caption{}
        \label{fig:velocity40}
    \end{subfigure}
    \hfill
    \begin{subfigure}[b]{0.45\textwidth}
        \includegraphics[ width=3in, height =1.60in]{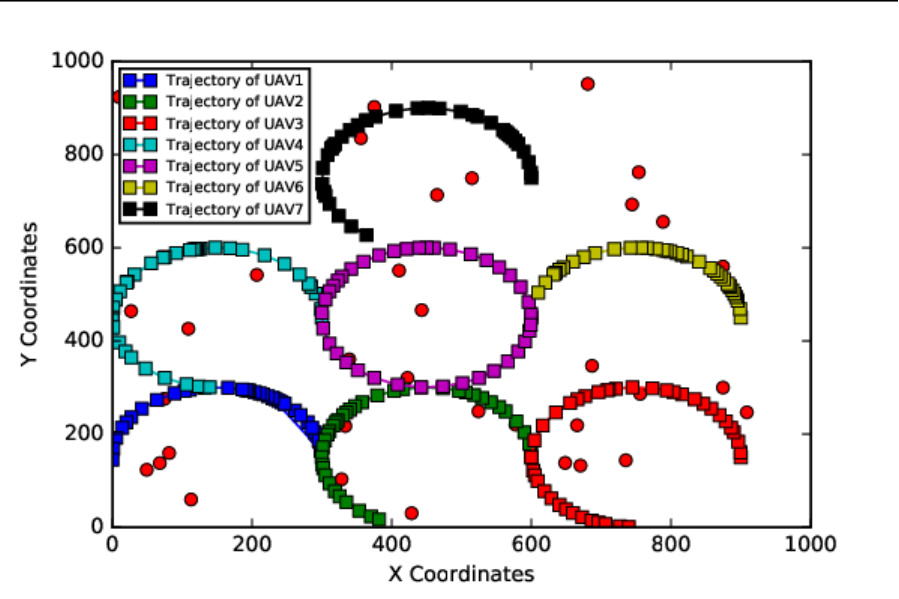}
        \caption{}
         \label{fig:trajectory40}
    \end{subfigure}
    \caption{Velocities and trajectories of MADRL-SA with I=7.(a) and (b)velocity and trajectory given number of waypoints=20. (c) and (d) velocity and trajectory given number of waypoints=40 }
    \label{fig:velfig}
\end{figure*}

\begin{figure}[ht] 
    	\includegraphics[ width=3.8in, height =2in]{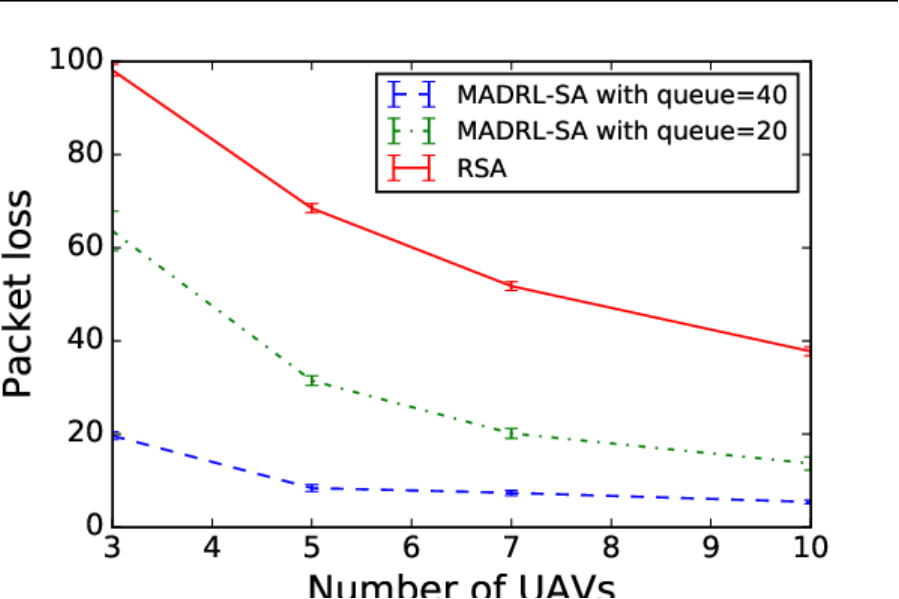}
            \Centering
    	\caption{Network cost with an increasing number of UAVs, where the data queue length of MADRL-SA is set to 20 and 40 and number of ground sensors as 40.}
    	\label{fig:increase}
\end{figure}

\begin{figure}[ht] 
    	\includegraphics[width=3.8in, height =2in]{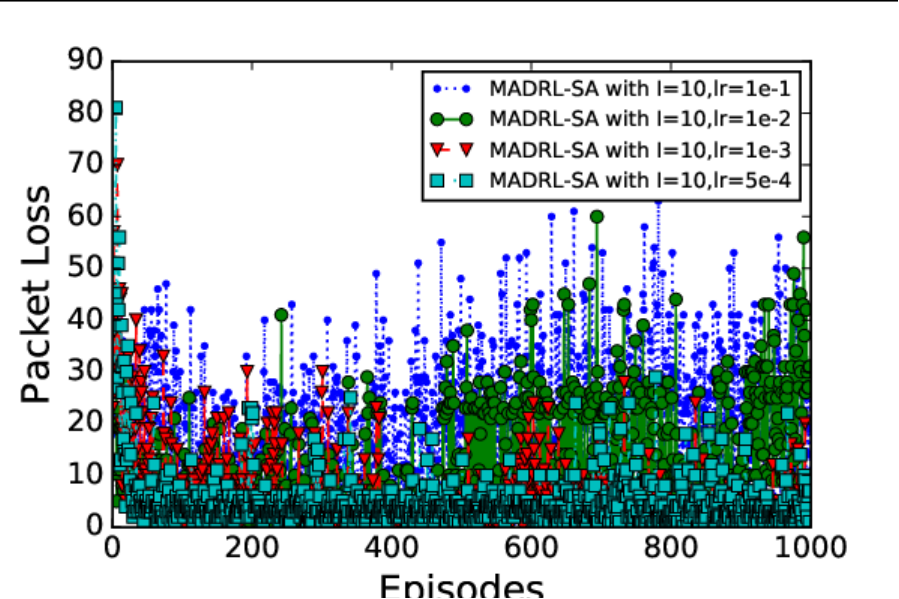}
            \Centering
    	\caption{Training performance with varied
learning rates.}
    	\label{fig:learning}
\end{figure}
Fig. \ref{fig:agent} depicts the convergence of MADRL-SA with I=10 for low and high SNR cases and DRL-SA. MADRL-SA with I=10 and high SNR show the best performance since it reduce the overflow cost as well as the fading cost due to good SNR. MADRL-SA with I=10 and low SNR outperform the DRL-SA which has the highest network cost. The reason is that when multiple UAVs act it results in the reduction of overflow cost. 

Fig. \ref{fig:network} depicts the network cost of MADRL-SA (data queue length=40) and the baselines in term of ground sensors. MADRL-SA with I=5 and I=10 achieves a lower network cost in comparison to CHSA. The network cost of MADRL-SA with I=5 is lower than that of  CHSA . Overall, MADRL-SA with I=5 and I=10 outperforms CHSA. Particularly, when J=100 the packet loss of MADRL-SA with I=5 and I=10 is lower than CHSA by around 21\% and 40\%, respectively.

Fig. \ref{fig:packet} shows the  trade-off between the number of ground sensors and UAVs. Specifically, a large number of ground sensors expedites the buffer overflows in UASNets and in turn, increases the packet loss. On the other hand, increasing the number of UAVs allows the ground sensors to be scheduled in parallel, hence reducing the buffer overflow. A balance needs to be struck between the numbers of UAVs and ground sensors  to minimize the packet loss.

Fig. \ref{fig:energy} shows the energy consumption of the ground sensors by varying the number of ground sensors and UAVs. For a given number of UAVs, the energy consumption of the network increases with the number of ground sensors. On the other hand, the increasing number of UAVs helps increase the number of ground sensors scheduled to transmit data, hence raising the energy consumption of the ground sensor network.

Fig. \ref{fig:velfig} show the velocities and trajectories of different UAVs for the MADRL-SA with I=7. Fig. \ref{fig:velfig}(a) demonstrates the velocity of 7 UAVs given 20 waypoints. The color bar shows the range of values for velocity and color map shows the actual velocity of each UAV for each waypoint in color format. As can be seen UAV 2 moves with the lowest velocity as confirmed by its small trajectory in Fig. \ref{fig:velfig}(b). In contrast, UAV 1 moves with the highest velocity as confirmed by its trajectory. Overall, for waypoints 1-12, UAV 3-7 move with the lowest velocity witnessing subtle changes. After these waypoints the velocity of these UAVs is increasing. 

Fig. \ref{fig:velfig}(c) is similar to Fig. \ref{fig:velfig}(a) except that number of waypoints is increased to 40. Overall, the pattern for all UAVs except UAV 5 is almost similar and all of them move with low or moderate velocity witnessing high velocity at some points, this can be confirmed by their associated trajectories in Fig. \ref{fig:velfig}(d). UAV 5 moves smoothly before waypoint 20. After this point its velocity start increasing and hence a full trajectory is shaped as can be seen in Fig. \ref{fig:velfig}(d).

Fig. \ref{fig:increase} evaluates the network cost with the increasing number of UAVs, where the buffer size of MADRL-SA is set to 20 or 40 and the number of ground sensors is 40. For MADRL-SA with buffer size of 40, increasing the number of UAVs from 3 to 10 leads to a reduction of the packet loss by 68\%. In contrast, when the buffer size is 20, a reduction of 77\% in the packet loss is witnessed. Fig. \ref{fig:increase} also shows that  MADRL-SA  significantly outperforms RSA by 80\% when the buffer size is 40, and by 34\% when the buffer size is 20. 

Fig. \ref{fig:learning} demonstrates the training performance with varied learning rates(lr). After few episodes in the beginning, the network cost have an obvious tendency to decrease and converge in the case of lr=1e-3 and lr=5e-4. Nevertheless, the algorithm may converge to a local optimum in case of large learning rate, this situation  can be seen in the case of lr=1e-1 and lr=1e-2.

\section{Summary}   \label{chap3:sec4}
In the first chapter, we study the joint flight cruise control and data collection scheduling in the UASNets. We formulate the problem using MMDP to minimize the packet loss due to buffer overflows at the ground sensors and fading airborne channels. We propose MADRL-SA to solve the formulated MMDP, where all UAVs utilize DQN to conduct respective decisions. In MADRL-SA, the UAVs acting as agents learn the underlying patterns of the data and energy arrivals at all the ground sensors as well as the scheduling decisions of the other UAVs. We conduct simulation using PyTorch deep learning library and results reveal that the proposed MADRL-SA for UASNets  reduces packet loss by up to 54\% and 46\%, as compared to the single agent case and  existing non-learning greedy algorithm, respectively. 
\section*{Appendix A}

The path loss of the LoS link is given by  
\begin{equation} \label{eq:40}
    PL_{LOS}=20\log d+20\log f+ 20 \log(\frac{4\pi}{c})+\eta_{LOS}
\end{equation}
The path loss of the non-LoS link is given by
\begin{equation} \label{eq:41}
    PL_{NLOS}=20\log d+20\log f+ 20 \log(\frac{4\pi}{c})+\eta_{NLOS}
\end{equation}

The LoS probability is given by   
\begin{equation} \label{eq:43}
    Pr_{LOS}=\frac{1}{1+a exp(-b[\varphi_j^i-a])}
\end{equation}

Then, the NLoS probability is

\begin{equation} \label{eq:42}
   Pr_{NLOS}=1-Pr_{LOS} 
\end{equation}

The expectation of the path loss  between  UAV \emph{$i$} and device \emph{$j$} can be obtained by 
\begin{equation} \label{eq:44}
\gamma_j^i=Pr_{LOS}\times PL_{LOS}+Pr_{NLOS} \times PL_{NLOS}  
\end{equation}

By substituting (\ref{eq:42}) into (\ref{eq:44}), we have
\begin{equation} \label{eq:45}
    \gamma_j^i=Pr_{LOS}(PL_{LOS}-PL_{NLOS})+PL_{NLOS} 
\end{equation} 
Substituting (\ref{eq:40}),(\ref{eq:41}),(\ref{eq:43}) into (\ref{eq:45}) leads to
\begin{multline} \label{eq:90}
\gamma_j^i=\frac{(\eta_{LOS}-\eta_{NLOS})}{1+a exp(-b[\varphi_j^i-a])}+20\log d+20\log f+\\ 20 \log(\frac{4\pi}{c})+\eta_{NLOS}
\end{multline} 

Rewriting \ref{eq:90} in term of $\varphi_j^i$ and r, we finally obtain
\begin{multline}
\gamma_j^i=\frac{(\eta_{LOS}-\eta_{NLOS})}{1+a exp(-b[\varphi_j^i-a])}+20 log(r \sec(\varphi_j^i))+20\log(\lambda)+\\ 20 \log(\frac{4\pi}{c})+\eta_{NLOS}
\end{multline}

\section*{Appendix B}
Let $\epsilon$ denote the bit error rate, $L$ denote the data packet length and $\lambda$ denote the packet arrival probability. Depending on the transmission status and arrival pattern, four transitions may happen as presented  in (\ref{eq:5}):
\begin{enumerate}
    \item In the first case, the packet transmission is successful $(1-\epsilon)^L$ and there is no packet arrival $(1-\lambda)$. The probability of such transition is $(1-\epsilon)^L \times (1-\lambda) $. Given L=R(t)*T where T is the conversation time of  UAV i and ground sensor j, and $T=\frac{2d_{i,j}}{v(t)}$. We have $L=\frac{2d_{i,j}R(t)}{v(t)}$ by substituting T into L. Therefore, the transition probability of the first case is  $(1-\epsilon)^{\frac{2d_{i,j}R(t)}{v(t)}}(1-\lambda)$.
    \item In the second case, the packet transmission is not successful $(1-(1-\epsilon)^L)$ and there is packet arrival $\lambda$. The probability of such transition is $(1-(1-\epsilon)^L) \times \lambda$. By substituting T into L, we have $L=\frac{2d_{i,j}R(t)}{v(t)}$. Therefore, the transition probability of the second case is  $(1-(1-\epsilon)^{\frac{2d_{i,j}R(t)}{v(t)}})\lambda$.
    \item  In the third case, the packet transmission is successful $(1-\epsilon)^L$ and there is packet arrival $\lambda$. The probability of such transition is $(1-\epsilon)^L  \times \lambda$. By substituting T into L, we have $L=\frac{2d_{i,j}R(t)}{v(t)}$. Therefore, the transition probability of the third case is $(1-\epsilon)^{\frac{2d_{i,j}R(t)}{v(t)}}\lambda$.
    \item In the fourth case, the packet transmission is not successful $(1-(1-\epsilon)^L)$ and there is no packet arrival $(1-\lambda)$. The probability of such transition is $(1-(1-\epsilon)^L)\times (1-\lambda)$. We have $L=\frac{2d_{i,j}R(t)}{v(t)}$. Therefore, the transition probability of the fourth case is  $(1-(1-\epsilon)^{\frac{2d_{i,j}R(t)}{v(t)}})(1-\lambda).$
\end{enumerate}

(\ref{eq:6}) investigates the transmission probabilities for unselected ground sensors. These ground sensors do not transmit data. In this case, the ground sensors either receive packet with transition probability $\lambda$ or no packet is received with transition probability $1-\lambda$.

           \clearemptydoublepage
	    \chapter{Age of Information Minimization using Multi-agent UAVs based on AI-Enhanced Mean Field Resource Allocation}
In this chapter, we introduce a cruise control approach based on MFG theory to minimize the AoI, while balancing the trade-off between UAVs' movements and AoI. This method reduces the complexity of the cruise control problem and enhances optimization of UAVs' movements. However, in practice, obtaining instantaneous knowledge of the UAV's cruise control decision and AoI is challenging, making the proposed MFG difficult to solve online. We formulate MMDP, with network states comprising the AoI of ground sensors and waypoints of the UAV swarm. The MMDP action space includes continuous waypoints and velocities, as well as discrete transmission schedules. We propose a mean field hybrid proximal policy optimization (MF-HPPO) approach. The rest of this chapter is organized as follows: In Section \ref{chap4:sec1}, we present the system model in which the channel model as well as the AoI in the UASNets is formulated. Moreover, we formulate the flight resource allocation of the UAV swarm as the MFG to minimize the AoI. Section \ref{chap4:sec2} develops the proposed MF-HPPO, to jointly optimize the cruise control of multiple UAVs and data collection scheduling. Section \ref{chap4:sec3} presents the implementation of the proposed MF-HPPO in Pytorch as well as performance evaluation. Finally, Section \ref{chap4:sec4} concludes this paper.

\section{Problem Statement} \label{chap4:sec1}
\subsection{System Model}  \label{sec:3}
\begin{figure} [ht]
        \centering
        \includegraphics[ width=3in, height =2in]{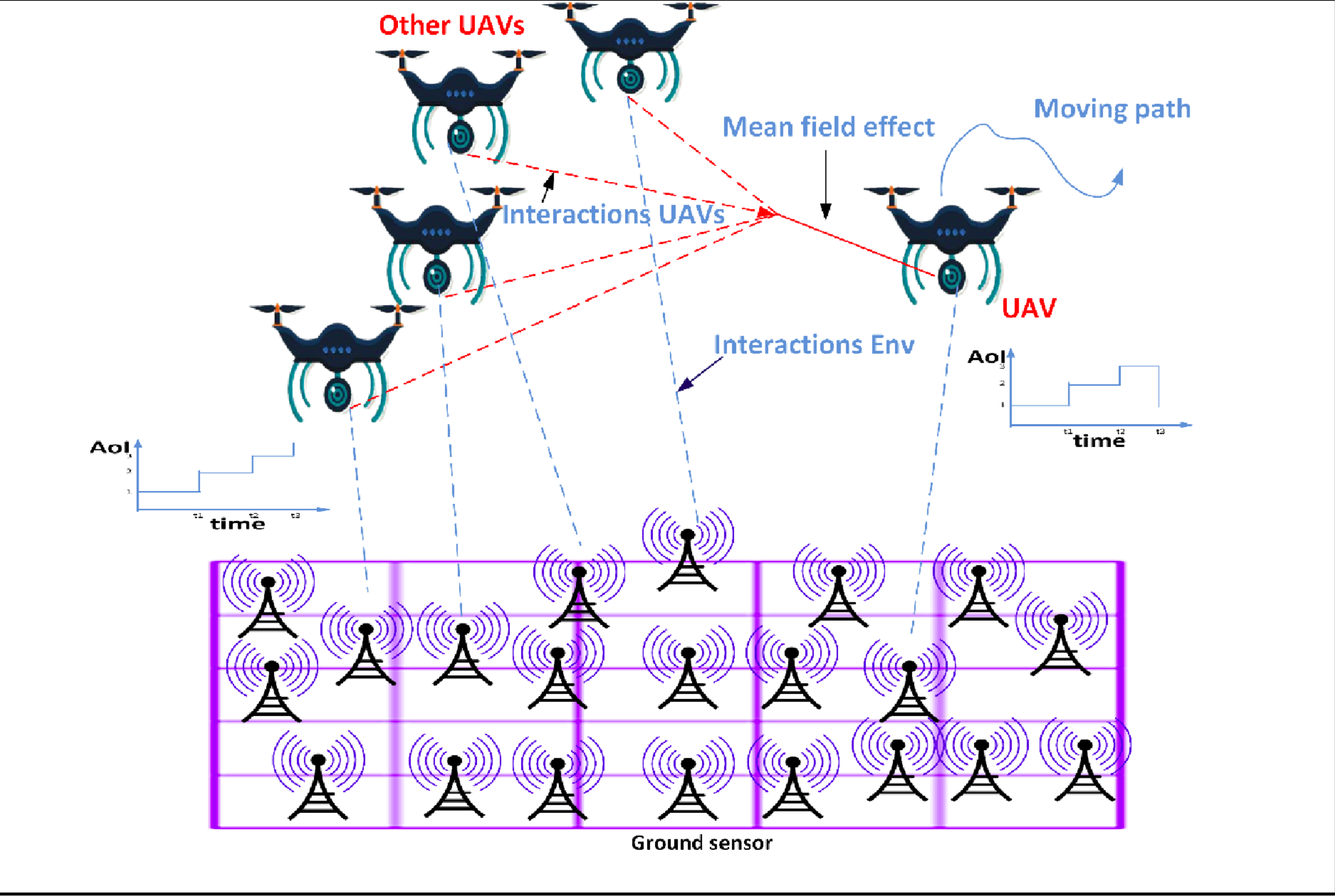}
    	\caption{Mean field representation of UASNets.}
    	\label{fig:forest}
\end{figure}

In this section, we present the system model of the considered UAVs-assisted sensor network. Notations used in this paper are summarized in Table \ref{table:100}. The system consists of \emph{$I$} UAVs, $\emph{$i$}\in[1,\emph{$I$}]$ and \emph{$J$} ground sensors, $\emph{$j$}\in[1,\emph{$J$}]$
in which the ground sensors are deployed in a target region. The UAVs are employed to patrol in the target zone while collecting the sensory data. Fig. \ref{fig:forest} depicts an example of UASNets along with mean field representation. With the increase in the number of UAVs in Fig. \ref{fig:forest} the interactions between them become complex and can dominate the overall behavior of the system. MFG designed to deal with the optimal control problem involving a large number of players. It has unique characteristics suitable for UAV swarm and modelling these interactions. Each UAV seeks to minimize the AoI according to the actions of other agents surrounded. As depicted, the UAV consider the mean field effect of the other UAVs, which represents the collective behavior of the UAVs in the system. The coordinates $(x_i,y_i,z_i)$ and $(x_j,y_j,0)$ represent the position of  UAV \emph{$i$} and ground sensor \emph{$j$}, respectively. The UAVs fly to the ground sensors, collect sensory data, and then their operation is terminated. The UAVs fly at a constant altitude, represented by $\zeta_i(t) = (x_i,y_i,z)$. The distance between ground sensor \emph{$j$} and UAV \emph{$i$} is $\sqrt{(x_i-x_j)^2+(y_i-y_j)^2+z^2}$. For the safety of the UAV during flight by preventing it from exceeding the maximum safe speed or stalling, we denote the maximum and
minimum velocity of the UAV as $v_{max}$ and $v_{min}$, respectively. 
\par
We consider that UAV \emph{$i$} moves in low attitude for data collection, where the probability of LoS communication between UAV \emph{$i$} and ground sensor \emph{$j$} is given by \cite{lap} 
\begin{equation} \label{eq:2}
    \Pr{_{LoS}}(\varphi_j^i)=\frac{1}{1+a \exp(-b[\varphi_j^i-a])}
\end{equation}
where \emph{$a$} and \emph{$b$} are constants, and  $\varphi_j^i$ denotes the elevation angle between  UAV \emph{$i$} and ground sensor
\emph{$j$}. Moreover, path loss of the channel between  UAV \emph{$i$} and device \emph{$j$} can be modeled by 
\begin{multline} \label{eq:3}
  \gamma_j^i=\Pr{_{LoS}}(\varphi_j^i)(\eta_{LoS}-\eta_{NLoS})+20 \log \left(r \sec(\varphi_j^i)\right)+ \\20 \log(\lambda)+20 \log \left(\frac{4\pi}{v_c}\right)+\eta_{NLoS}    
\end{multline}
where \emph{$r$} is the radius of the radio coverage of UAV \emph{$i$}, \emph{$\lambda$} is the carrier frequency, and \emph{$v_c$} is the speed of light. $\eta_{LoS}$ and $\eta_{NLoS}$ are the excessive path losses of LoS or non-LoS, respectively.

\begin{table}[ht]
\centering
\caption{Notation and Definition}
\begin{tabular}{|c|c|} 
 \hline
 \textbf{Notation} & \textbf{Definition}\\  
 \hline
 \emph{$J$} & number of ground sensors  \\ 
 \hline
 \emph{$I$} & number of UAVs \\
 \hline
  $h_j^i(t)$  & channel gain between device j and UAV i \\
 \hline
 $\zeta_i(t)$  & location of the UAV on its trajectory\\
 \hline 
 $v_i(t)$  & velocity of UAV i\\
 \hline
 $v_{max},v_{min} $  & the maximum and minimum velocity of UAV i\\
 \hline
 $M$ & number of episodes \\
 \hline
 $L$ & length of each episode \\
 \hline
 $\gamma$ & discount factor \\
 \hline
 $\eta$ &  learning rate \\ 
 \hline
 $D$ & buffer size \\
 \hline
 $B$ & mini-batch size \\
 \hline
 $a_i$ & action of UAV $i$ \\
 \hline
 $o_i$ & mean field of UAV $i$ \\
 \hline
 $a_i^c$ & continuous action of UAV $i$ \\
 \hline
 $a_i^d$ & discrete action of UAV $i$\\
 \hline
 $s_{\alpha,i}$ & state of UAV $i$\\
 \hline
 $E[..]$ & mathematical expectation\\
 \hline
 $A$  & advantage function \\
 \hline
 $\theta$ & network parameter \\
 \hline
 $\pi$ & policy \\
 \hline
 $\pi^c$ & continuous policy \\
 \hline
 $\pi^d$ & discrete policy \\
 \hline
 $\sigma$ & diffusion coefficient  \\
 \hline
 $W$ & weiner process \\
 \hline
 $H$ & entropy \\
 \hline
 \end{tabular}
\label{table:100}
\end{table}  
To characterize the freshness of the collected sensory data at the UAV, AoI is defined as the time that has passed since ground sensor generates the latest information. The AoI of ground sensor $j$ that generated a data packet at $t_j$ and collected by UAV $i$ at $t_i$ is given by
\begin{equation} \label{eq:333}
AoI^i_j(t)=t_i-t_j.
\end{equation}
\noindent According to (\ref{eq:333}), it can be also known that maintaining a low $AoI^i_j(t)$ is critical for improving the effectiveness and timeliness of the sensory data, reducing the response time, and providing real-time information for decision-making at the UAVs. 
\subsection{Problem Formulation} \label{sec:4}
In this section, we formulate the MFG optimization with a large number of UAVs to address the trade-off between the cruise control of the UAVs and AoI. We also explore the FPK equation to determine the optimal velocities of the UAVs while characterizing the collective behavior of the UAVs. We begin with optimal control formulation in Section \ref{sub:41} and then proceed with MFG formulation in Section \ref{sub:42}.
\subsubsection{Optimal Control Formulation} \label{sub:41}
We derive the state dynamics and cost function, then we formulate the velocity control problem using the optimal control theory.
\begin{enumerate}
    \item Time-varying Dynamics of Network States:
Let $\zeta_i(t)$ denote the position of the UAV $i$ at time $t$ and $v_i(t)$ denotes the velocity. 
According to Newton's laws of motion \cite{waldrip2013explaining}, the location dynamics of UAV $i$ can be expressed by 
\begin{equation} \label{eq:4}
    d\zeta_i(t)=v_i(t) dt+ \sigma dW_i(t)
\end{equation}
where $W_i(t)$ is a standard Wiener process \cite{morters2010brownian} with a diffusion coefficient $\sigma$. 
\item Cost Function:
Each UAV intends to optimize its velocity to minimize the cost function. Our cost is defined as the average AoI of all ground sensors.
The average AoI can be computed as:
\begin{equation} \label{eq:7}
    c(t)=\frac{1}{IJ}\Sigma_{j=1}^J \Sigma_{i=1}^I AoI^i_j(t). 
\end{equation}
\item Velocity Control Problem Formulation:
Given a period of time $T$ regarding the data collection, the velocity of UAV $i$ at $t$, denoted as $v_i^*(t)$, is optimally controlled to minimize $c(t)$, which gives:
\begin{equation} \label{eq:8}
v_i^*(t)= \underset{v_i(t)}{\arg \min} E\left[\int_0^T c(t)dt)\right],
\end{equation}
\[ s.t.\ (\ref{eq:4}).\]
\end{enumerate}

To determine $v_i^*(t)$ in (\ref{eq:8}), classical game theories, such as differential game, fails to capture the aggregate behavior of all the UAVs. Differential game assumes each agent's movement is independent of others. This assumption fails to capture the fact that a large number of UAVs' trajectories decisions are influenced by the aggregate behavior of all the UAVs, thus hardly minimizing the average AoI, $c(t)$. 

We novelly extend MFG to capture the impact of the aggregate behavior of the UAVs, in terms of cruise control. The MFG models the aggregate decision of UAVs as a probability distribution, rather than focusing on the actions of individual UAVs. This recognizes that the cruise control of each UAV is influenced by the behavior of all other UAVs. Moreover, the formulated MFG is defined to minimize $c(t)$ given a large number of UAVs, which classical game theory struggles with due to the computational complexity of solving for the equilibrium. 
\subsubsection{MFG Problem Formulation} \label{sub:42}
We reformulate the optimal cruise control problem in (\ref{eq:8}) into a cooperative MFG problem. The computational complexity of the system is greatly reduced by formulating an MFG, since a large number of interactions with other agents is converted into an interaction with the mass.
The interaction between each UAV with the other UAVs is modeled as a mean-field term, which is denoted by $m(\zeta(t))$. The mean-field term is the distribution over agents´ state space or control to model the overall state and control of them. We can measure the state and control of all agents in an MFG using the mean-field term.\\

Given dynamics, $\zeta_i(t)$, the mean-field term of $m(\zeta(t))$ can be denoted by
\begin{equation} \label{eq:9}
 m(\zeta(t))=\lim_{I\to\infty} \frac{1}{I} \Sigma_{i=1}^{I} \mathbbm{1} \{\zeta_i(t)=\zeta(t)\},   
\end{equation}
where $\mathbbm{1}$ is an indicator function which returns 1 if the given
condition is true, or 0, otherwise.

Given $m(\zeta(t))$, the state dynamics, cost function and FPK equation can be defined as:\\
\begin{itemize}
    \item  \textbf{State dynamics:} The state dynamics of each UAV  can be expressed by

       \begin{equation} \label{eq:10}
            d\zeta(t)=v(t) dt+ \sigma dW(t).
       \end{equation}
    \item  \textbf{Cost function:} The mean-field term affects the running cost function of each UAV. The average AoI of the all UAVs is computed by
\begin{equation} \label{eq:11}
    c(v(t),m(\zeta(t)))=\int c(v(t))\cdot m(\zeta(t)) d\zeta.
\end{equation}
Mathematically, the cost function can be written by
\begin{equation} \label{eq:12}
    J(v(t),m(\zeta(t))))=\int_{t=0}^T c(v(t),m(\zeta(t)) dt.
\end{equation}
If the UAV move quickly, lead to poor channel condition and retransmissions thereby AoI prolongs. In contrast, slow movement of the UAV, may prolong the AoI of the ground sensors because the data are not collected in time. The cost function addresses these trade-offs and find the optimal velocity to balance these objectives.
\item  \textbf{Focker-Planck equation:} Based on (\ref{eq:10}) we develop the FPK equation. The FPK equation governs the evolution of the mean field function of UAVs and given by:
\begin{equation} \label{eq:13}
    \partial_tm(\zeta(t))+\nabla_{\zeta}m(\zeta(t))\cdot v(t)- \frac{\sigma^2}{2}\nabla_{\zeta}^2m(\zeta(t))=0. 
\end{equation}
See \it{Appendix}.
\end{itemize}
After deriving the state dynamics, cost function, and FPK equation, we now proceed to present the MFG. 

To summarize, the cooperative MFG problem is given by
\begin{equation} \label{eq:14}
\min_{v,m} J(v(t),m(\zeta(t)))
\end{equation}
\[ s.t.\ (\ref{eq:13}).\]
\section{Proposal} \label{chap4:sec2}
\subsection{Proposed MF-HPPO} \label{sec:5}
In this section, we describe the MFG as an MMDP in Section \ref{sub:52} so that the optimal actions of UAVs can be learned by the proposed MF-HPPO. MF-HPPO is presented in Section \ref{sub:53}, which employs onboard PPO to minimize the average AoI  of the ground sensors. The trajectory and instantaneous speed of the UAVs, and the selection of the ground sensors are optimized in a mixed action space. In Section \ref{sub:54}, an LSTM  layer is developed with MF-HPPO  to capture the long-term dependency of data.
 
\subsubsection{MMDP Formulation} \label{sub:52}
We reformulate the MFG using MMDP framework to enable the application of PPO for optimizing the actions and minimizing average AoI. By adapting the MMDP framework to our problem, we define the relevant state space, action space, transition probabilities, policy and cost function, thus facilitating an effective solution approach based on MF-HPPO. We define our MMDP as follows.
\begin{itemize}
    \item {\em Agents}: the number of agents, i.e., UAVs is denoted by \emph{I}.
    \item {\em State}: A state $s_{\alpha}$ of the MMDP consists of the positions of UAV $i$, the AoI of ground sensors, i.e, $s_{\alpha}$=\{$\zeta_i(t)$,$AoI^i_j(t):i \in [1,I],j \in [1,J] $\}. All states of the MMDP constitute the state space.
    \item {\em Action}: Each UAV $i$ takes an action $a_i$ that schedules a ground sensor for data transmission and determines the flight trajectory and velocity, i.e, $a_i$ =\{$k_j^i$,$v_i(t)$,$\zeta_i(t)$\}
    \item {\em Policy}: Policy $\pi_i$ is the probability of taking each action of agent i.
    \item {\em State Transition}: The current state $s_{\alpha}$ transit to a new state $s_{\beta}$ according to probability $P(s_{\beta}\mid s_{\alpha},a)$, where $a$ indicates a joint action set that includes the actions of all the UAVs.
    \item {\em Cost}: The immediate cost of the UAVs is $\frac{1}{IJ}\Sigma_{j=1}^J \Sigma_{i=1}^I AoI^i_j(s_{\alpha},a)$.    
   
\end{itemize}
\subsubsection{MF-HPPO} \label{sub:53}
The proposed MF-HPPO operates onboard at the UAVs to determine their trajectories and  sensor selection. The UAV chooses a sensor and moves to it, then sends out a short beacon message with the ID of the chosen sensor. Upon the receipt of the beacon message, the selected sensor transmits its data packets to the UAV, along with  the state information of $AoI^i_j(t)$ in the control segment of the data packet. After the UAV correctly receives the data, it sends an acknowledgement to the ground sensor.

The following equation highlights the mean field idea of MF-HPPO \cite{yang2018mean}:
\begin{multline} \label{36}
    Q_{i}(s_{\alpha,i},a)=\frac{1}{N_i}\Sigma_{k\in N(i)} Q_i(s_{\alpha,i},a_i,a_k)=Q_i(s_{\alpha,i},a_i,o_i).
\end{multline}
Here, $Q_i$ is the $Q$ value of agent $i$, $a$ represents the joint action of all agents. The neighbor agents of agent $i$ are characterized by $N_i$. $o_i$ is an indicator of the mean field.
In essence, in multi-agent systems the Q value of an agent is computed based on the current state and joint action, but when we have a large number of agents computing joint action is impractical, therefore (\ref{36}) allow an agent to compute its Q value just based on the mean field of its neighbors. 

Fig. 2 shows the proposed MF-HPPO with LSTM layer, where each UAV equipped with the MF-HPPO to minimize the average AoI by optimizing the trajectory and data collection schedule.
The use of the LSTM layer, continuous and discrete actors, and the objective function of PPO, are the features of the MF-HPPO in this diagram. As shown, The decision-making component of each agent consists of two actors and a critic, which is preceded by the LSTM layer to draw conclusions based on experience. The actor for continuous action spaces outputs continuous values for cruise control, such as position and velocity, and the actor for discrete action spaces outputs a categorical value that can be used to select one of the ground sensors. Each agent samples the actions and performs in the environment. The rollout buffer is filled with data generated by these interactions such as, state, mean field, action, cost and policy. As can be seen, we use  Generalized Advantage Estimate (GAE) \cite{schulman2015high} as a sample-efficient method to estimate the advantage function. As depicted, based on the RolloutBuffer, mini-batches are then formed to train the LSTM and the actors and critics so that the agent can continuously improve its policies. The definition of the objective function of PPO is the total of actor losses and critic loss subtracted by entropy, as depicted in the diagram. The actor loss is inputted by the ratio of old policy and current policy and the advantage value. The critic loss is inputted by the critic's output and the return value. The policy is designed to encourage the agent to take advantageous actions, while punishing actions that deviate from the current policy.
\par
Algorithm \ref{al:1} summarizes the MF-HPPO with the LSTM-based
characterization layer. In the initialization step, Input and Output are characterized; the algorithm receives parameters like Clip threshold, discount factor and mini-batch size as input and specify its output as trajectory and scheduling policy of UAV $i$. Next, the actor $\pi_i$  and critic $w_i$ are initialized with random weights for each agent. The number of training episodes is $M$, where the length of each episode is $L$.
Each agent is trained using a predetermined set of iterations throughout the learning phase. Sampling and optimization constitutes the learning phase. In the beginning of learning, the state $s_{\alpha,i}$ and mean field $o_i$ are randomly  initialized for each agent. With the start of the sampling policy, UAV $i$ samples its action based on the policy $\theta_{old}^i$. The sampled action represents sensor selection, velocity and locations, and executed in the environment to obtain the cost, new state and new mean field. Consequently, trajectories (i.e., sequence of states, actions, policy, mean field, and costs) are gathered and stored in the RolloutBuffer. In addition, GAE is applied to calculate the advantage that is used in (\ref{eq:37}). In the optimization step, the policies are optimized. In the optimization step, the policy parameter is updated for each epoch. The PPO objective is computed in each epoch according to the following equation:
\begin{equation} \label{eq:37}
          \begin{split}
             & L^{clip}(\theta^i)=\min\Bigl(\frac{\pi_{\theta^i}(a_i|s_{\alpha,i},o_i)}{\pi_{\theta_{old}^i}(a_i|s_{\alpha,i},o_i)}A_{\pi_{\theta_{old}^i}}(s_{\alpha,i},o_i,a_i),
             \\
             & g(\epsilon,A_{\pi_{\theta_{old}^i}}(s_{\alpha,i},o_i,a_i))\Bigr)
          \end{split}
\end{equation}
where
\begin{equation} \label{eq:100}
    \pi_{\theta^i}(a_i|s_{\alpha,i},o_i)=\pi_{\theta^i}^c(a_i^c|s_{\alpha,i},o_i)\pi_{\theta^i}^d(a_i^d|s_{\alpha,i},o_i).
\end{equation}
\begin{algorithm} [ht] 
\caption{MF-HPPO Characterized by LSTM Layer}  \label{al:1}
\textbf{1.Initialize}
\\
\KwIn {Clip threshold $\epsilon$, discount factor $\gamma$, learning rate $\eta$, buffer size $D$, mini-batch size $B$ }
\KwOut {The scheduled ground sensor j and trajectory $\zeta_i$ of UAV i}
Randomly initialize the Actors $\pi_i$ and Critics $w_i$ with networks parameters $\theta^i$
\\
The LSTM layer with $\{W_o, W_c, W_f, W_p\}$ and $\{e_o, e_c, e_f, e_p\}$.
\\
Initialize the sampling policy $\pi_{\theta_{old}^i}$ with
$\theta_{old}^i = \theta^i$.
\\
$\forall i \in (1,I)$
\\
\textbf{2.Learning}
\\
\For {$episode=1$ \KwTo $M$}{
      Randomly obtain the initial state $s_{\alpha,i}$ 
      \\
      \For {$t=1$ \KwTo $L$}{
       \hspace{2cm}\textbf{*The sampling phase*}
       \\
       Sample: Sample action $a_i \sim \pi_{\theta_{old}^i}(a_i|s_{\alpha,i},o_i,\theta^i)$;
       \\
       Execute the action $a_i$ that specifies
       the scheduled ground sensor j  and trajectory $\zeta_i$ of  UAV i.
       \\
       Obtain the cost and new state $s_{\beta, i}$ and new mean field $o_i(t+1)$.
       \\
       RolloutBuffer: store the trajectory $( s_{\alpha,i},a_i,c,o_i,\pi_{\theta_{old}^i}(a_i|s_{\alpha,i},o_i,\theta^i))$
       \\
       $s_{\alpha,i}=s_{\beta,i}$
                           }
         Compute the advantage using GAE
         
         \For {$epoch=1$ \KwTo $P$}{
         
          \begin{flushright}
       \end{flushright}  
         \hspace{2cm}\textbf{*The optimization phase*}
       
         Sample the RolloutBuffer
         \\
         Compute the PPO-Clip objective function using ($\ref{eq:37}$) 
         \\
         Compute the critic loss.
         \\
         Optimize the overall objective function using ($\ref{eq:38}$) 
          }
     Synchronize the sampling policy
     $\pi_{\theta_{old}^i}  = \pi_{\theta^i}$ 
     \\
    Drop the stored data in RolloutBuffer.
   }
\end{algorithm}
Here $a_i^c$ and $a_i^d$ correspond to actions in continuous and discrete spaces. In (\ref{eq:100}), to obtain the hybrid policy $\pi_{\theta^i}(a_i|s_{\alpha,i},o_i)$, we multiply the policies for continuous and discrete actions \cite{neunert2020continuous}. Meanwhile, we assume that wireless radio of the UAV can cover the whole field.

Continuous policy $\pi_{\theta^i}^c$ is modeled using multivariate normal distribution and discrete policy $\pi_{\theta^i}^d$ is modeled using categorical distribution. 
In the next step, the overall objective function is optimized according to the following equation:

\begin{equation} \label{eq:38}
          L^{total}(\theta^i)=L^{clip}(\theta^i)-K_1 L^{VF}(\theta^i)+K_2*H.
\end{equation}
\begin{figure*}[ht] 
        \centering
        \captionsetup{justification=centering}
    	\includegraphics[ width=5in]{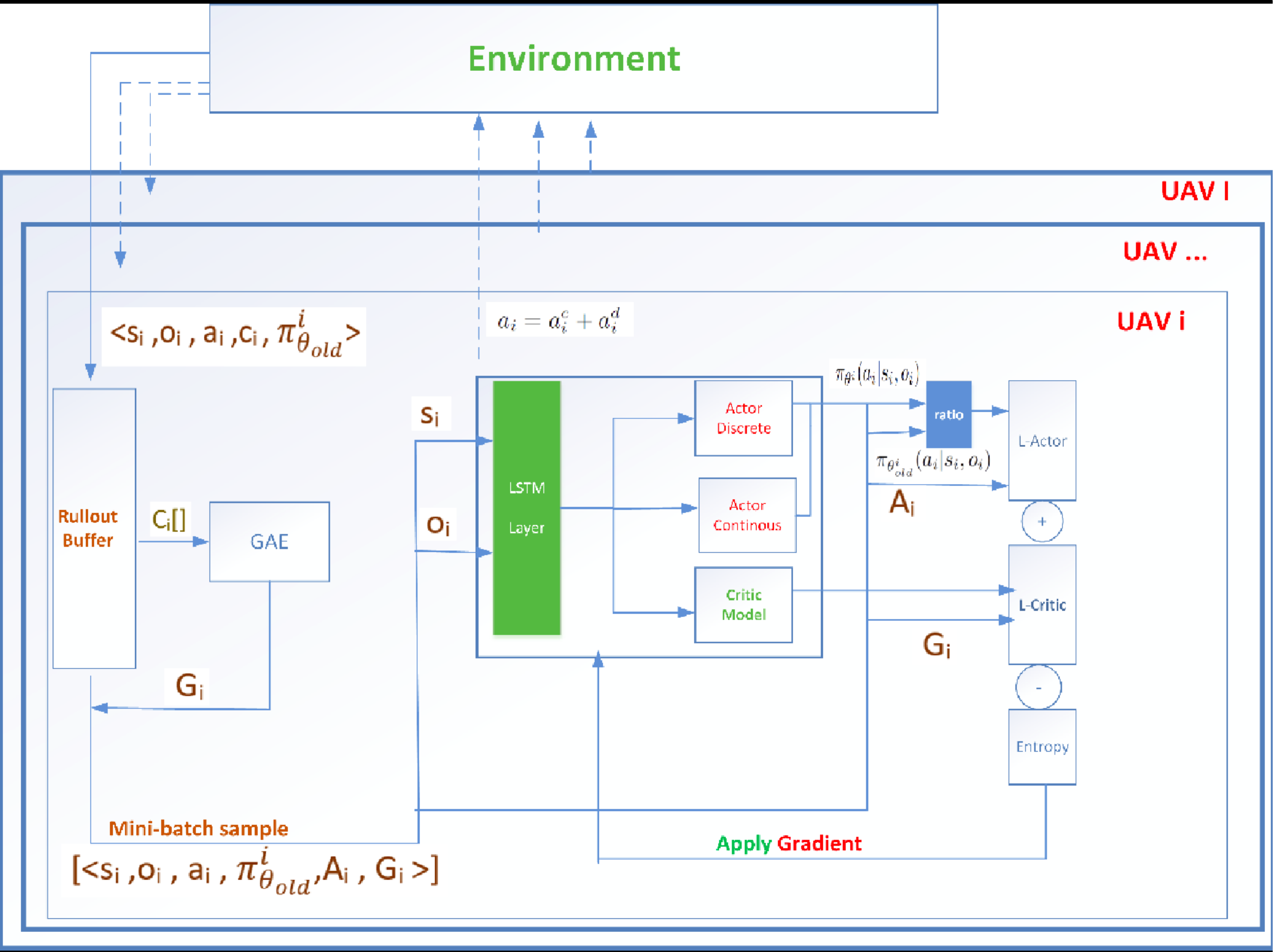}
    	\caption{Overview of MF-HPPO: Each UAV equipped with LSTM layer to optimize 
discrete and continuous actions using hybrid policy
. 
}
    	\label{fig:MF-HPPO}
\end{figure*}

\noindent Here, $L^{VF}(\theta^i)$ is the critic loss and $H$ acts as a regularizer encourages the agent to execute actions more unpredictably for exploration and guard against the policy being overly deterministic. The entropy for continuous and discrete actions is computed based on the actions' distribution. We obtain the entropy by  multiplication of the entropy of continuous and discrete action spaces to enable enforcing consistent regularization to both continuous and  discrete action spaces. $K_1$ balances the importance of the critic loss and the actor loss, and $K_2$ coefficient controls the amount of entropy in the policy. 

Finally, the sampling policy $\pi_{\theta_{old}^i}$ is updated with the policy $\pi_{\theta^i}$, and the
stored data are dropped. The next iteration then begins.
\subsubsection{LSTM Layer} \label{sub:54}
We further develop an LSTM layer in the proposed MF-HPPO, which captures long-term dependencies of time-varying network state $s_\alpha$. Cell memory and the gating mechanism are main components of LSTM. Cell memory is responsible to store the summary of the past input data and the gating mechanism regulates the  information flow between the input, output, and cell memory. The network states are fed into LSTM one by one (one at each step). The last hidden state $\kappa_i^{hidd}$  is returned as the output of the state characterization layer.
Each agent uses an LSTM layer to predict their
respective hidden states. The hidden states $\kappa_i^{hidd}$
 are calculated by the following
composite function:
\begin{equation}
    \kappa_i^{hidd}=out_itanh(C_i),
\end{equation}
\begin{equation}
    out_i=\sigma(W_0 \cdot [C_i,\kappa_{i-1}^{hidd},A_i]+e_i ),
\end{equation}
\begin{equation}
    C_i=F_iC_{i-1]}+p_itanh(W_c.[\kappa_{i-1}^{hidd},A_i]+e_c),
\end{equation}
\begin{equation}
    F_i=\sigma(W_f \cdot [\kappa_{i-1}^{hidd},C_{i-1},A_i]+e_f),
\end{equation}
\begin{equation}
     p_i=\sigma(W_f \cdot [\kappa_{i-1}^{hidd},C_{i-1},A_i]+e_p),
\end{equation}
where the output gate, cell activation vectors, forget gate, and input gate of the LSTM layer are denoted by $out_i$, $C_i$, $F_i$, and $p_i$, respectively. $\sigma$ and tanh correspond to logistic sigmoid
function and the hyperbolic tangent function, respectively.
${W_0,W_c,W_f ,Wp}$ are the weight matrix, and
${e_0, e_c, e_f , e_p}$ are the bias matrix\cite{9507550}, \cite{9779339}.
\subsection{Complexity and Convergence of MF-HPPO}
The overall complexity of MF-HPPO is calculated as follows,
$O(I \cdot ML \cdot(\Sigma_{g=1}^G n_{g-1}.n_g))$
where $n_g$ is the number of neural units in the g-th hidden layer. In this work, the PPO architecture is built with the same $n_g$ in all hidden layers. Therefore, the PPO complexity can be reduced to $O(I \cdot ML \cdot(g-1) \cdot n_g^2)=O(I \cdot ML \cdot n_g^2)$. The convergence analysis is proved by simulation results (see Fig. \ref{fig:conv}).
\section{Evaluation} \label{chap4:sec3}
\subsection{Implementation of MF-HPPO}
MF-HPPO is implemented in Python 3.8 using Pytorch (the Python deep learning library). A Predator Workstation running 64-bit Ubuntu 20.04 LTS, with Intel Core i7-11370 H CPU @ 3.30 GHz 8 and 16 GB memory is used for the Pytorch setup. Table  \ref{table:2} clearly outlines the
different considered simulation parameters.
MF-HPPO algorithm is trained over 3000 episodes with 40 steps each. The discount factor and learning rate are set
to 0.99 and 3e-4, respectively. Each agent comprises the input layer, LSTM layer, the critic and actors with fully-connected hidden layers of size 256 and output layer. Each neuron uses Rectified Linear Unit (ReLU) as an activation function. In addition, Hyperbolic tangent (tanh) and softmax are used as activation functions in the output layer of the continuous actor-network and discrete actor network. The input of each critic network is represented as a concatenation of states and mean field, and its output is a scalar that assesses the states according to the global policy. The total log probability of the hybrid policy is the sum of the log probabilities of the continuous and discrete action spaces. This log probability would be used as part of the calculation of the objective function in MF-HPPO, along with the estimated cost and the entropy regularization term.
\subsection{Baseline Description}
The MF-HPPO characterized with LSTM layer is compared by single-agent PPO, random scheduling and trajectory design (RSTD), multi-agent DQN (MADQN) and MF-HPPO without LSTM Layer.
\begin{figure*}[ht]
    \centering
  
    \begin{subfigure}[b]{0.45\textwidth}
        \centering
        \includegraphics[ width=\textwidth]{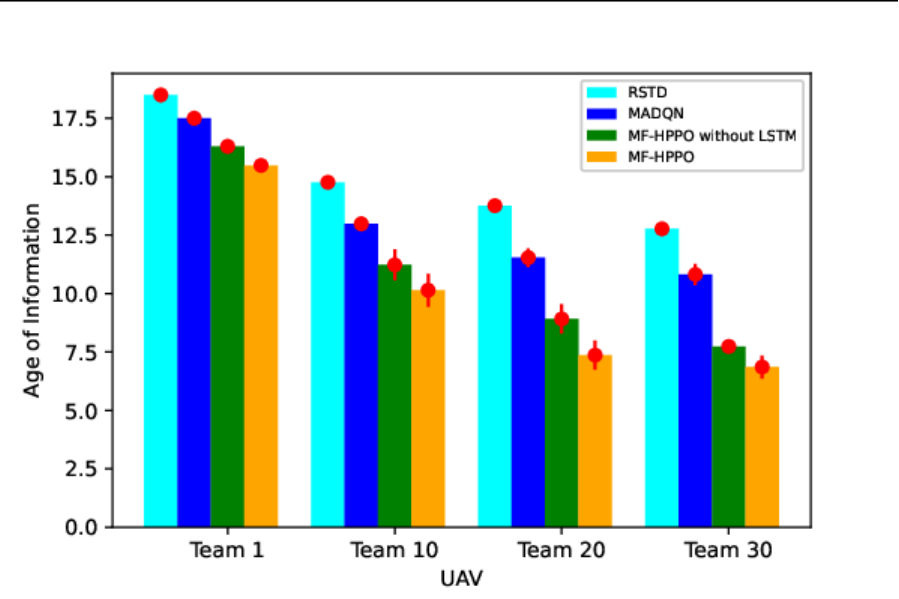}
        \caption{Evaluation of MF-HPPO's performance with a variable number of UAVs in comparison to RSTD, MADQN and MF-HPPO without LSTM}
        \label{fig:multiuavn}
    \end{subfigure}
    \hfill
    \begin{subfigure}[b]{0.45\textwidth}
        \centering
        \includegraphics[ width=\textwidth]{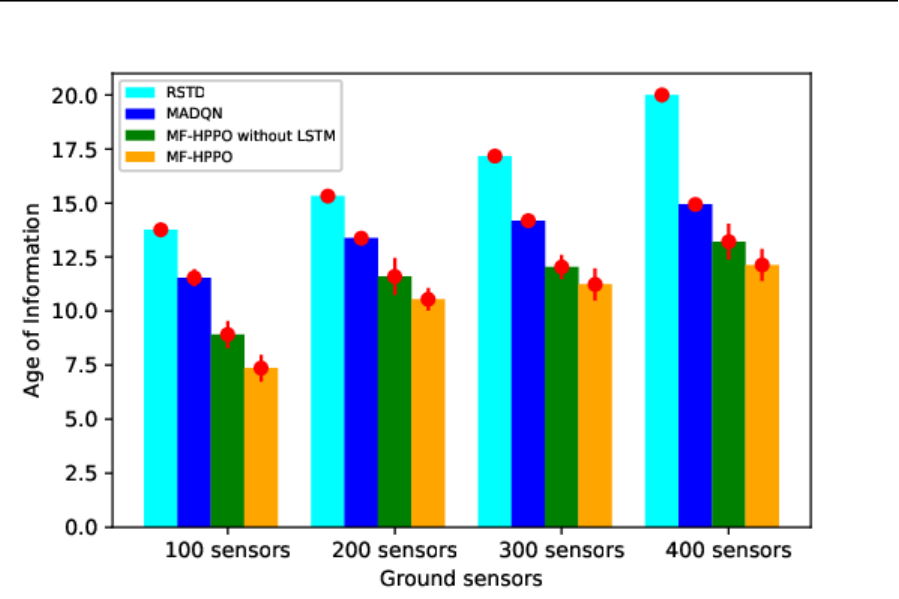}
        \caption{Evaluation of MF-HPPO's performance with a variable number of ground sensors in comparison to RSTD, MADQN and MF-HPPO without LSTM}
        \label{fig:ground sensorsn}
    \end{subfigure}

      \caption{Performance evaluation of MFFPO by changing the number of UAVs and ground sensors }
    \label{fig:uavsensor}
\end{figure*}
\begin{figure} [ht!]
        \includegraphics[width=3.5in, height =2.5in]{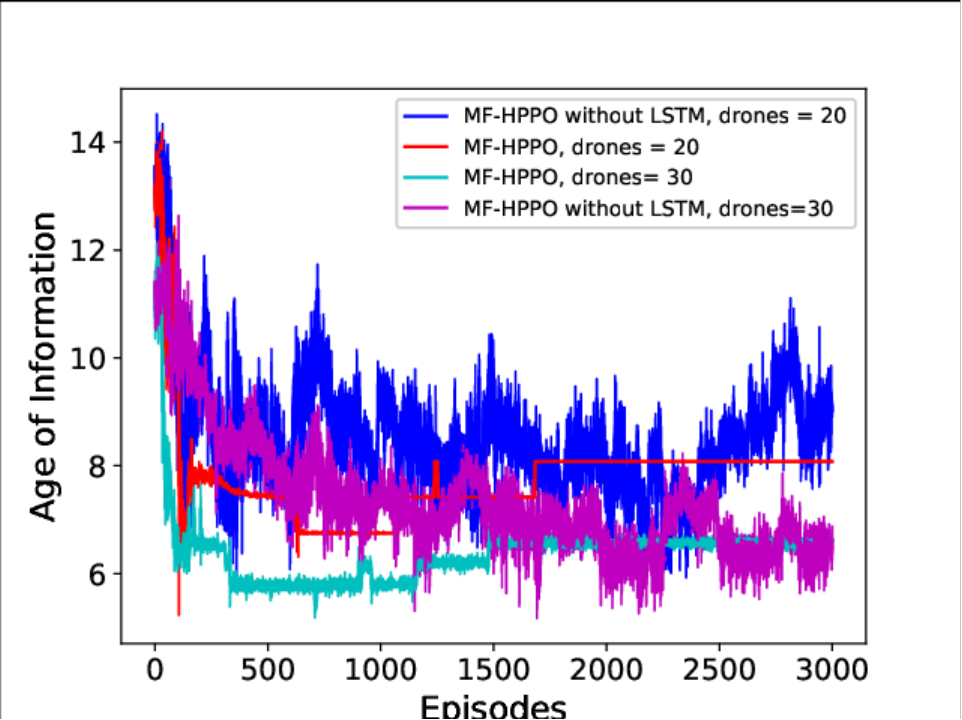}
            \Centering
    	\caption{The network cost for each episode of MF-HPPO with I=30 and benchmarks}
    	\label{fig:conv}
\end{figure}
A brief introduction of the four benchmarks is given below
\begin{enumerate}
    \item PPO, in this algorithm single-agent running PPO to optimize trajectory and transmission scheduling.
    \item RSTD, in this algorithm transmission scheduling and trajectory design, are randomly designed. 
    \item MADQN, in this algorithm, each agent running DQN cooperate to reduce average AoI following circular trajectories.
    \item MF-HPPO without LSTM Layer, the structure of this algorithm is same as MF-HPPO but without LSTM layer.
\end{enumerate}
\begin{table}[ht]
\centering
\caption{PyTorch Configuration}
\begin{tabular}{|c|c|} 
 \hline
  \textbf{Parameters} &  \textbf{Values} \\
 \hline
 Number of ground sensors & 100\\ 
 \hline
 Number of UAVs & 30 \\
 \hline
 Geographical area size [m] & 1,000*1,000 \\
 \hline
 Altitude of the UAVs & 120 m \\
 \hline
  Activation Function for Hidden Layers & Relu  \\
 \hline
 Activation Function for Continuous Action & Tanh \\
 \hline
 Activation Function for Discrete Action & Softmax \\ 
 \hline
 Critic Network Learning Rate & 3e-4 \\
 \hline
 Actor Network Learning Rate & 3e-4 \\
 \hline
 Number of Hidden Layers for Networks & 2 \\
 \hline
 Number of Neurons & 256  \\
 \hline
 Loss Coefficients for $K_1$ and $K_2$ & 0.2 and 3 \\
 \hline
 Optimizer Technique & Adam \\
  \hline
  Clip Fraction & 0.2 \\
  \hline
 Rollout Buffer size & 40 \\
 \hline
 Batch size & 40 \\
 \hline
 Mini Batch Size & 4 \\
 \hline
 PPO Epochs & 8 \\
 \hline
  Number of episodes & 3,000 \\
 \hline
 Discount Factor & 0.99 \\
 \hline
\end{tabular}
\label{table:2}
\end{table}
\subsection{Performance analysis of MF-HPPO }
Fig. \ref{fig:uavsensor} depicts the performance evaluation of MF-HPPO in comparison to the baselines by changing the number of UAVs and ground sensors.
Fig. \ref{fig:multiuavn} shows the impact of the number of UAVs on the AoI. Overall, the AoI decreases when more UAVs are deployed because time efficiency increases and more ground sensors can be operated in less time. Increasing the number of UAVs from 1 to 30 result in a 61\% decrease in the average AoI for MF-HPPO, while that of MADQN is 37\%. The reason is that MF-HPPO performs the optimization in a mixed action space with higher training stability than MADQN with circular trajectories.
Fig. \ref{fig:ground sensorsn} evaluates the average AoI given 20 UAVs and groups of 100, 200, 300, and 400 ground sensors. The MADQN and the RSTD are used as baselines. Overall, increasing the number of ground sensors results in a uniform increase in the average AoI, since more sensor data should be collected. In particular, when the number of ground sensors is 400, the proposed MF-HPPO outperforms the RSTD by 38\% and the MADQN by 17\%.

We obtain the convergence trend of MF-HPPO in Fig. \ref{fig:conv} by deploying 20 UAVs serving 100 ground sensors. In general, the proposed MF-HPPO (I=30) achieves the lowest AoI compared to MF-HPPO without LSTM layer (I= 20 and 30) with a gain of 33\% and 66\%, respectively. Since the trajectories and scheduling of data collection for multiple UAVs are optimized with better time efficiency. At the same time, the LSTM layer enables better exploration as agents use experience to guide their actions. Moreover, thanks to the LSTM layer, convergence is accelerated and stabilized. The peak AoI of the proposed MF-HPPO drops significantly from 14 seconds to 6 seconds in the first 1,000 episodes. From episode 1,500 to episode 3,000, the AoI stabilizes at 7 seconds with minimal fluctuations.
\begin{figure*}[ht]
    \centering
  
    \begin{subfigure}[b]{0.32\textwidth}
        \centering
        \includegraphics[ width=\linewidth]{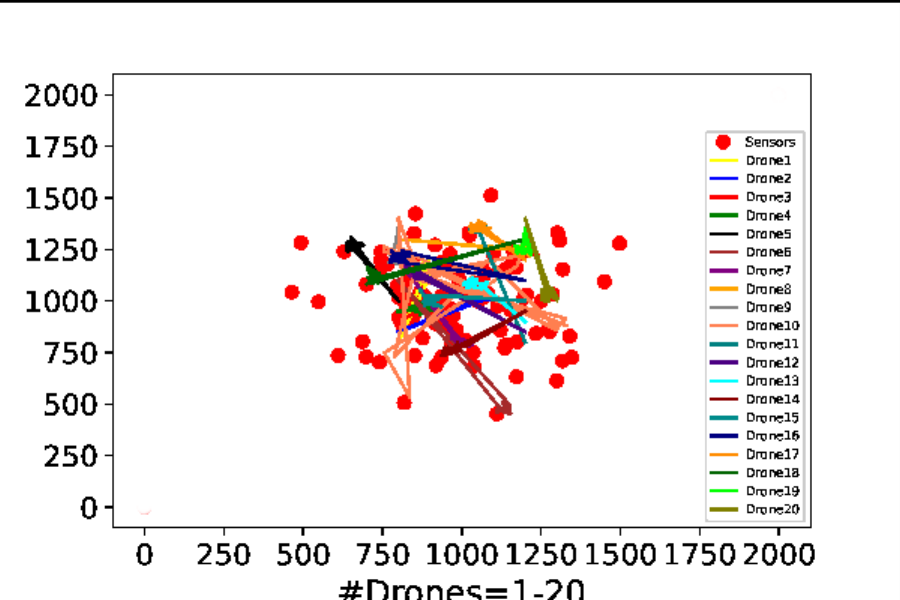}
        \caption{Normal Distribution}
        \label{fig:normal}
    \end{subfigure}
    \hfill
    \begin{subfigure}[b]{0.32\textwidth}
        \centering
        \includegraphics[ width=\linewidth]{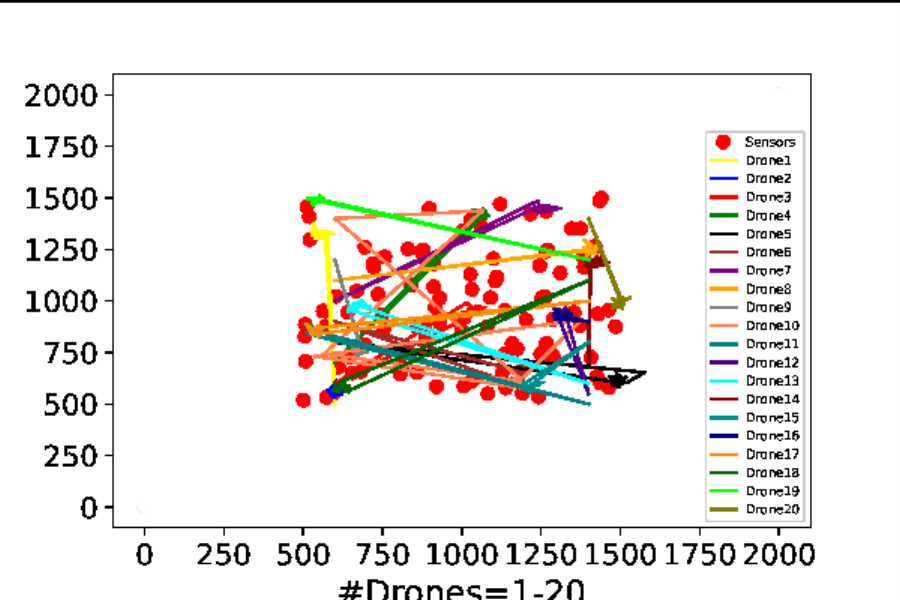}
        \caption{Square Distribution}
        \label{fig:square}
    \end{subfigure}
    \hfill
    \begin{subfigure}[b]{0.32\textwidth}
        \centering
       
        \includegraphics[width=\linewidth]{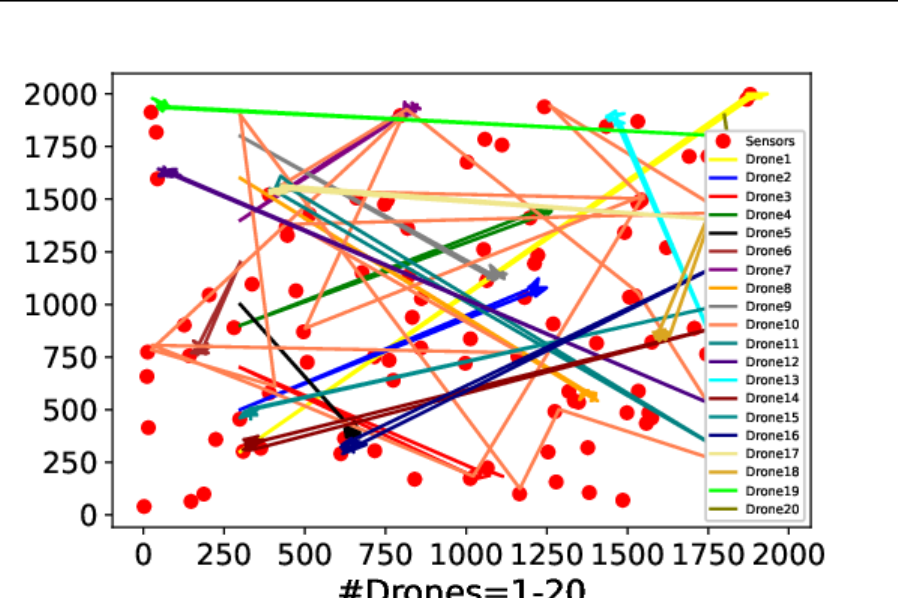}
        \caption{Uniform Distribution}
        \label{fig:uniform}
    \end{subfigure}

    \caption{MF-HPPO trajectory distributions for various UAV counts and ground sensor distributions. }
    \label{fig:four figures}
\end{figure*}

MF-HPPO-generated trajectories for 20 UAVs are shown in Fig. \ref{fig:four figures}, where the ground sensor distribution patterns are uniform, square, or normal ones.
When designing trajectories for AoI minimization, the UAVs' trajectories are impacted by the distribution of the ground sensors. The UAV needs to approach to the location of each scheduled sensor to collect the data and update its AoI.
Fig. 5(a), refer to the normal distribution and shows  trajectories for 20 UAVs, focusing on the center area of the ground sensors and less on the corners. The normal distribution of the ground sensors can affect the UAVs' trajectories by determining which ground sensors are prioritized for data collection. For example, as can be seen, most ground sensors are centered and their data may become stale, in this case, the UAVs'trajectories are designed to visit these ground sensors more frequently to minimize the average AoI. Figs. 5(b) is related to the square distribution. As can be seen, the ground sensors are less centered. This cause diverse set of ground sensors in wider range to be covered in comparison to normal distribution. Fig. 5(c) refer to the uniform distribution. As can be seen, the UAVs design wide-area trajectories due to the wider distribution of ground sensors covering the entire area and the AoI requirements of the scattered ground sensors.

\begin{figure} [ht]
        \includegraphics[width=3.5in, height =2.5in]{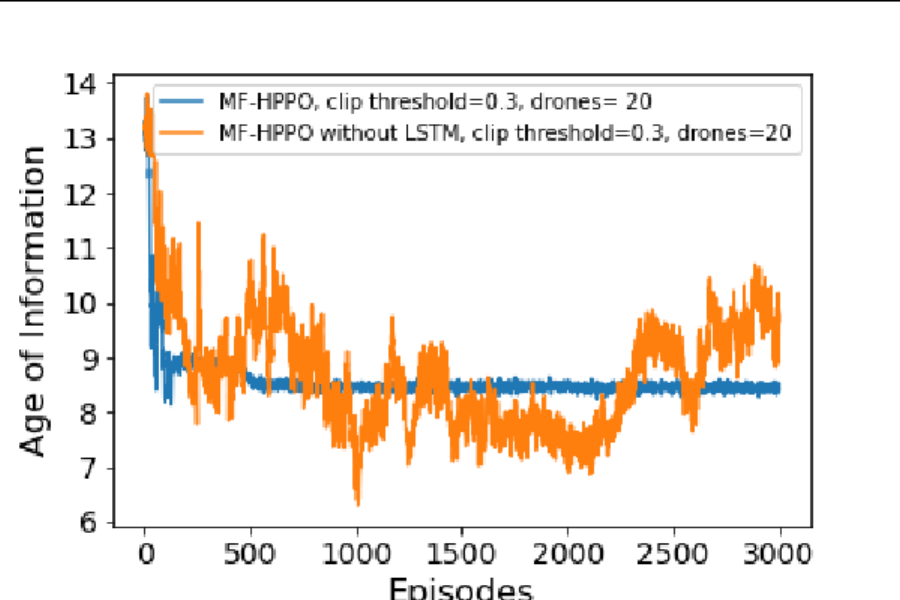}
            \Centering
    	\caption{Performance evaluation of MF-HPPO by changing clip threshold}
    	\label{fig:clip}
\end{figure}

Fig. \ref{fig:clip} demonstrates the convergence figures for two variants of MF-HPPO by changing the clip threshold.
PPO uses the clip threshold, commonly referred to as epsilon, to regulate the amount of policy updating. A larger clip threshold allows 

\noindent for more aggressive updating, while a smaller clip threshold restricts updating more severely, resulting in less policy change. The blue curve shows the MF-HPPO with LSTM layer and a clip threshold of 0.3 outperforming the MF-HPPO without LSTM layer clip threshold 0.3. The latter shows a deviating behavior due to the influence of the clip threshold, while the blue curve shows an absolutely stable trend despite the same value of the clip threshold thanks to the LSTM layer. Overall, adding the LSTM layer to MF-HPPO can stabilize the training and prevent divergence of the strategies.
\section{Summary} \label{chap4:sec4}
In this chapter, we propose a mean field flight resource allocation to model velocity control for a swarm of UAVs, in which each UAV minimizes the average AoI by considering the collective behavior of others. Due to the high computational complexity of MFG, we  leverage AI and propose MF-HPPO characterized with an LSTM layer to optimize the UAV trajectories and data collection scheduling in mixed action space. Simulation results based on PyTorch deep learning library show that the proposed MF-HPPO for UASNets reduces average AoI by up to 57\% and 45\%, as compared to existing non-learning random algorithm and MADQN method (which performs the action of trajectory planning in the discrete space), respectively. This confirms the AI-enhanced mean field resource allocation is a practical solution for minimizing AoI in UAV swarms.

\section*{Proof of FPK Equation for Cruise Control}
We derive the mean field via an arbitrary test function $g(\zeta)$, which is a twice continuously differentiable compactly supported function of the state space. The integral of $m(\zeta)g(\zeta)d\zeta$ can be considered as the continuum limit of the sum $g(\zeta(t))$, where $\zeta(t)$ is the UAV's state at time $t$. It is known that,

\begin{equation}  \label{eq:410}
    \int m(\zeta(t)) g(\zeta) d\zeta=\frac{1}{N}\Sigma_{i=1}^N g(\zeta(t)).
\end{equation}
At time $t$, the first-order differential function with regard to time $t$ is derived to check how this integral varies in time. By utilizing the chain rule, we can derive the heuristic formula as	
\begin{multline} \label{eq:420}
    \int \partial_t m(\zeta(t)) g(\zeta) d\zeta=\\ \frac{1}{N}\Sigma_{i=1}^N \partial_t \zeta(t)\nabla g(\zeta(t))+\partial_t^2 \zeta(t) \nabla^2 g(\zeta(t)).
\end{multline}
Taking the limit of the right side of the above equation when $N$ tends
to infinity, we get 
\begin{multline}  \label{eq:430}
    \int [ \partial_tm(\zeta(t))+\nabla_{\zeta}m(\zeta(t))\cdot \frac{\partial \zeta}{\partial t}- \\ \frac{\eta^2}{2}\nabla_{\zeta}^2m(\zeta(t))] g(\zeta(t))d\zeta=0,
\end{multline}
for any test function $g$ through integration by parts. Then the above equation leads to the following equation:
\begin{equation} \label{eq:440}
    \partial_tm(\zeta(t))+\nabla_{\zeta}m(\zeta(t))\cdot v(t)- \frac{\sigma^2}{2}\nabla_{\zeta}^2m(\zeta(t))=0. 
\end{equation}
which correspond to FPK equation defined in (\ref{eq:13}).
            \clearemptydoublepage
            \chapter{Conclusions and Future Work}
\label{chap:Conc}
\section{Summary}
We employed UAVs for data collection from ground sensors in harsh environments, such as crop monitoring. The use of UAVs for data collection offers advantages such as improved network throughput and extended coverage range beyond terrestrial gateways. However, a major challenge arises from the impact of UAV movements on channel conditions, leading to packet loss or outdated packets.
To address this challenge, we proposed a joint optimization approach to minimize packet loss by controlling the velocities of multiple UAVs and optimizing their data collection schedules. Our proposed solution, MADRL-SA, enables UAVs to asymptotically minimize packet loss even when they have outdated knowledge of the network states.
Furthermore, we introduced a novel mean-field flight resource allocation optimization method to minimize the AoI for sensory data. This involved formulating the trade-off between UAV movements and AoI as an MFG. To tackle practical scenarios, we proposed the MF-HPPO scheme, which optimizes UAV trajectories and data collection scheduling for ground sensors using a combination of continuous and discrete actions. Additionally, we incorporated LSTM to predict the time-varying network state and enhance training stability in MF-HPPO.
We conducted extensive simulations to evaluate the effectiveness of our proposed approaches. The results demonstrated that MADRL-SA reduced packet loss by up to 54\% and 46\% compared to existing solutions involving single UAV with DRL and non-learning greedy heuristics, respectively. Similarly, the simulation results showed that MF-HPPO reduced the average AoI by up to 45\% and 57\% compared to the MADQN method and non-learning random algorithm, respectively.

\section{Future Works}
MADRL-SA and MF-HPPO can be enhanced with explainable AI and human-in-the-loop mechanisms to significantly improve their effectiveness and usability. By enriching these approaches, we can leverage human expertise, provide transparent explanations for decisions, enhance performance, foster user trust, and promote better collaboration between humans and AI systems in UASNets. These enhancements have the potential to improve the efficiency and reliability of communication scheduling and cruise control, ultimately enhancing the overall operation of UASNets.
One approach to achieve this is by using feature importance techniques to identify and quantify the contribution of input features to the decisions made by DRL algorithms. By developing visualizations that depict the relationships between input features, intermediate algorithm states, and output decisions, we can provide users with a better understanding of the model’s decision-making process.
Another avenue is to incorporate human feedback to shape the cost function utilized by DRL algorithms. By shaping the cost function based on human preferences, we can guide the algorithms to make decisions that align better with human values and expectations. 	Furthermore, adopting an interactive ML paradigm allows human experts to interact with DRL algorithms during the training process. Human feedback, in the form of instructions or corrections, can be integrated into the learning process to improve the model’s performance.
In conclusion, the integration of explainable AI and human-in-the-loop reinforcement learning in this thesis can significantly enhance the performance and usability of the proposed algorithms in UASNets.

            \clearemptydoublepage






	\bookmarksetup{startatroot}								
	\addtocontents{toc}{\bigskip}							
	\renewcommand{\bibname}{Bibliography}					
	\phantomsection
	\addcontentsline{toc}{chapter}{Bibliography}			
	\begin{singlespace}
		\bstctlcite{IEEEexample:BSTcontrol}					
		\nocite{*}											
		\bibliographystyle{IEEEtranSN}						
		\bibliography{References}							
	\end{singlespace}
	\clearemptydoublepage



\end{document}